\documentclass[12pt]{article}

\usepackage{setspace}
\usepackage[english]{babel}

\usepackage[letterpaper,top=1in,bottom=1in,left=1in,right=1in,marginparwidth=1in]{geometry}

\usepackage[sort]{natbib}
 \bibpunct[, ]{(}{)}{,}{a}{}{,}%

\usepackage{amsmath}
\usepackage{graphicx}
\usepackage{graphicx}
\graphicspath{ {figures/} }
\usepackage[table]{xcolor}
\usepackage{tabularx}
\usepackage{booktabs,xltabular}
\usepackage{longtable}
\usepackage{afterpage}
\usepackage[title,titletoc]{appendix}
\usepackage{caption}
\usepackage{subcaption}
\usepackage{multicol}
\usepackage{multirow}

\usepackage[colorlinks=true, allcolors=blue]{hyperref}
\usepackage{csquotes}

\usepackage{color}
\definecolor{clemson-orange}{RGB}{234,106,32}
\definecolor{broncos-orange}{RGB}{252,76,2}
\definecolor{cincinnati-red}{RGB}{190,0,0}
\definecolor{chicago-maroon}{RGB}{128,0,0}
\definecolor{northwestern-purple}{RGB}{82,0,99}
\definecolor{pink}{RGB}{255,105,180}
\definecolor{celtics}{RGB}{46,123,59}
\definecolor{leafs-blue}{RGB}{0,58,120}
\definecolor{pure-cyan}{RGB}{0,100,92}
\definecolor{lawngreen}{RGB}{0,250,154}

\usepackage{tcolorbox}
\usepackage{csquotes}
\tcbset{
  promptbox/.style={
    colback=gray!5,
    colframe=gray!50,
    boxrule=0.5pt,
    arc=2pt,
    left=6pt,
    right=6pt,
    top=6pt,
    bottom=6pt
  }
}
\usepackage{makecell}

\providecommand{\keywords}[1]
{
  \small	
  \textbf{Keywords:} #1
}

\newtheorem{hypothesis}{Hypothesis}
\usepackage{authblk}
\usepackage{framed}

\title{From Dyads to Groups: Rethinking Emotional Support with Conversational AI}
\date{\vspace{-5ex}}

\author{
    \large{Yuqing Hu}\\ 
    \small{Michael G. Foster School of Business, University of Washington, \\ yuqinghu@uw.edu} 
    \and 
    \large{Wendao Xue} \\
    \small{McCombs School of Business, The University of Texas at Austin, \\ Wendao.Xue@mccombs.utexas.edu} 
    \and
    \large{Yifan Yu} \\
    \small{McCombs School of Business, The University of Texas at Austin, \\ Yifan.Yu@mccombs.utexas.edu} 
    \and
    \large{Yong Tan} \\
    \small{Michael G. Foster School of Business, University of Washington, \\ ytan@uw.edu} \\
}
\begin{document}
\maketitle
\newpage
\begin{abstract}
    Advances in artificial intelligence (AI), together with persistent gaps in access to reliable emotional support, have positioned AI as an increasingly prominent source of emotional assistance. However, most AI-based emotional support applications and prior research focus on one-on-one interactions between users and a single AI agent, leaving the potential advantages of alternative support configurations largely unexplored. Drawing on social support and support group theory, this research examines whether AI-based emotional support delivered by a group of AI agents (group AI support) can constitute a more effective support form than single-agent support (single AI support). We propose that group AI support enhances users’ perceived support efficacy, that this effect operates by strengthening users’ connectedness with the AI system, and that the composition of support types within AI groups further shapes support outcomes. Three experiments provide convergent support for these claims. By identifying when and why group AI emotional support outperforms single AI support, this work advances theoretical understanding of AI-based emotional support and provides actionable guidance for the design of AI support systems.
\end{abstract}

\keywords{Emotional Artificial Intelligence, Conversational Agent, Emotional Support, Social Support, Support Group Theory}

\section{Introduction}\label{sec:intro}

Emotional support plays a critical role in helping individuals cope with stress, uncertainty, and adverse life events. Yet access to timely, sustained emotional support remains uneven and constrained \citep{americanpsychologicalassociationAPAPollReveals, dawesProjectedCostsEconomic2024}. Growing emotional support needs, increasing social isolation, and barriers to traditional support resources have led many individuals to seek alternative sources of emotional assistance \citep{americanpsychologicalassociationHealthAdvisoryUse2025, wangRelationshipsSocialSupport2025, welchDigitalInterventionsReduce2023}. In parallel, advances in artificial intelligence (AI) have substantially expanded its capacity to engage in emotionally relevant interactions, including recognizing affect, responding with empathy, and sustaining supportive dialogue \citep{sorinLargeLanguageModels2024, ovsyannikovaThirdpartyEvaluatorsPerceive2025, inzlichtPraiseEmpathicAI2024}. As a result, AI-based emotional support systems are increasingly deployed in everyday contexts to provide companionship, reassurance, and coping guidance. Emotional support and companionship have been identified as the most common reasons people use generative AI tools \citep{zao-sandersHowPeopleAre2025}. Between 2022 and mid-2025, the number of AI companion applications increased by 700\% \citep{efuaandohAIChatbotsDigital2026, perezAICompanionApps2025}. Together, these developments position AI as an emerging participant in the emotional support landscape. 

Despite this rapid growth, most AI-based emotional support applications adopt a one-on-one interaction form, in which users interact with a single AI agent at a time. We refer to this configuration as single AI support. In contrast, emotional support in human settings typically takes more varied forms, as individuals are embedded in social networks and often receive support from multiple others simultaneously. For example, people frequently share concerns in group chats, receive feedback from several friends on social media posts, or turn to online forums and peer support groups when personal networks provide limited assistance. A natural way to reflect these social dynamics in AI systems is to create a support group composed of multiple AI agents, which we refer to as group AI support. In line with this idea, group forms of AI emotional support are beginning to emerge in practice. For example, an application named Duxiang, which translates to “solo echo” in English, allows users to post thoughts and receive responses from multiple AI characters acting as friends. These observations highlight an underexplored form of AI-based emotional support and point to new opportunities for exploring alternative designs that enhance user experience and support effectiveness.

Although intuitively appealing, whether group AI represents a more effective form of emotional support remains an open question in existing research. Scholars have increasingly recognized the capability of AI as a source of emotional support. They have devoted substantial attention to developing AI systems for emotional support and empirically examining their effectiveness in supporting users’ emotions \citep{liuEmotionalSupportDialog2021, qianHarnessingPowerLarge2023, liSystematicReviewMetaanalysis2023a}. However, consistent with prevailing business practices, this literature has focused predominantly on emotional support delivered through a single interacting AI agent (e.g., a chatbot). Consequently, the existing AI support literature offers little direct evidence on the impact of group AI support. Moreover, findings from human group support contexts cannot be assumed to generalize automatically to AI support groups. AI agents differ fundamentally from human supporters in identity, agency, and motivation \citep{kyungRationallyTrustEmotionally2025, lugerHavingReallyBad2016, houDoubleEdgedRolesGenerative2025}, which may shape how support is delivered, interpreted, and evaluated. At the same time, AI enables configurations of group support that are difficult or sometimes impossible to implement in human groups, such as flexible group size, rapid response timing, and controlled composition of support types. Together, these considerations underscore the theoretical and empirical value of examining whether, how, and under what boundary conditions group AI support may outperform single AI support.

We seek to advance the literature by asking the following research question: Can group AI support constitute a more effective form of emotional support than single AI support? If so, what psychological mechanisms underlie this advantage, and which configurations of AI groups enhance support effectiveness? Drawing on support group theory in the social support literature \citep{cohenStressSocialSupport1985,langfordSocialSupportConceptual1997,cohenSocialSupportMeasurement2000b}, this study develops a theoretical account of how and why group-based AI emotional support may differ from single-agent support. We focus on users’ perceived support efficacy, a core construct in social support and psychological research that reflects individuals’ subjective evaluations of how helpful support is for managing distressors \citep{pauwAvatarWillSee2022a, riniEffectiveSocialSupport2006, aroraPerceivedHelpfulnessImpact2007, kimPursuitComfortPursuit2006, smithClientPerceptionsTherapy2013, cocklinClientPerceptionsHelpfulness2017}. 

Support group theory suggests that human support from groups can outperform one-on-one support through three pathways: a cognitive pathway (e.g., offering a broader range of experiential knowledge and coping strategies), an affective pathway (e.g., a warm environment that promotes feelings of being understood), and a social pathway (e.g., fostering a temporary community that enhances perceived social support) \citep{cohenSocialSupportMeasurement2000b}. While it is relatively uncontroversial that groups of AI agents could provide emotional support through the cognitive and affective pathways \citep{yinAICanHelp2024a,bilquiseEmotionallyIntelligentChatbots2022}, it is less obvious whether an AI group offers advantages over a single AI agent through the social pathway. This motivates us to explore whether AI group support, which provides multiple supportive ties, fosters a stronger sense of connection than a single AI agent. Specifically, we argue that the group form enhances recipients’ connectedness with the AI support system. Connectedness refers to recipients’ active involvement in a meaningful relationship with another entity \citep{townsendConnectednessReviewLiterature2005}. Heightened connectedness, in turn, amplifies the influence of supportive messages and leads to higher evaluations of support efficacy. 

In addition to increasing the number of agents from one to multiple -- a change that primarily reflects the \textit{structural} support dimension -- group AI also raises questions about the \textit{functional} support dimension \citep{langfordSocialSupportConceptual1997,stewartFunctionalStructuralSocial2022a}. It is unclear how the functional composition of support should be configured within the group. More specifically, emotion-focused support, such as expressions of care and empathy, and information-focused support, such as advice or guidance, represent the most common and relevant functional categories \citep{cohenSocialSupportHealth1985,yanFeelingBlueGo2014a,liuEmotionalSupportDialog2021}. \footnote{The two support functions are more commonly referred to as emotional and informational support in prior research. We use the terms emotion-focused and information-focused to avoid conflating emotional support as a specific support function with emotional support as the broader goal of alleviating an individual’s negative emotions, given that both concepts are used in this paper.}

Building on these arguments, we propose that group AI emotional support leads to higher perceived support efficacy than single-agent AI support. We further propose that this effect is mediated by users’ perceived connectedness with the AI system. Finally, we propose that the functional composition of emotion-focused and information-focused support within AI groups further shapes perceived support efficacy.

\begin{figure}[!htbp]
    \centering
    \includegraphics[width=0.8\textwidth]{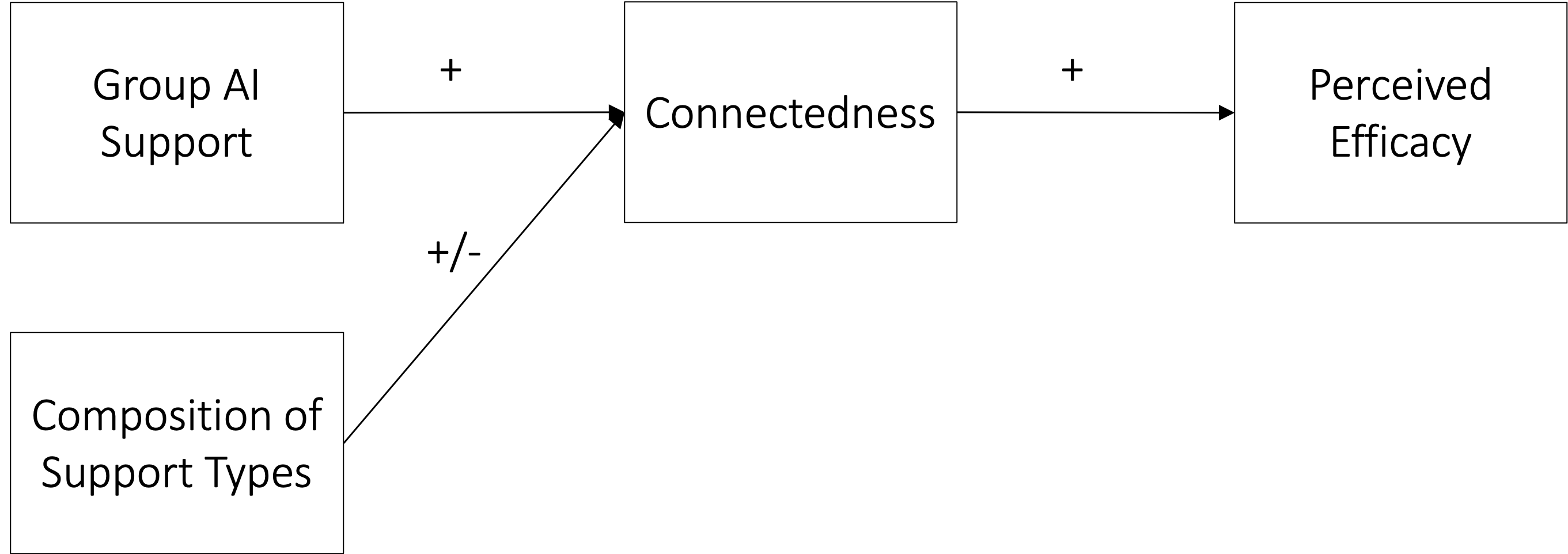}
    \caption{Theoretical Framework}
    \label{fig:framework}
\end{figure}

To test these hypotheses, we conducted three experiments in which participants received emotional support from either a group of AI agents or a single AI agent through a web-based experimental system developed for this study. The results support our hypotheses. In particular, AI group support improves perceived support efficacy through increased connectedness. 
Such an effect is relatively insensitive to variation in group size (i.e., the structural dimension), whereas the functional composition of AI agents meaningfully shapes support outcomes. 

Our findings provide three primary contributions. First, this study advances the literature on AI-mediated emotional support by identifying group AI support as a distinct and more effective support form. Drawing on social support and support group theories \citep{cohenStressSocialSupport1985,langfordSocialSupportConceptual1997,cohenSocialSupportMeasurement2000b}, we redirect attention from dyadic, one-on-one support provided by one AI agent to the broader configuration of the AI support environment. Second, by identifying connectedness \citep{townsendConnectednessReviewLiterature2005} as a mediating mechanism, we highlight the importance of relational experience in AI-based emotional support. This is a nascent area in the AI-based emotional support literature because it remains unclear whether and how users form relational bonds with non-human agents \citep{skjuveLongitudinalStudyHuman2022}. Third, the findings also delineate the boundary conditions for the efficacy of group AI support. Our results suggest that simply increasing structural support in AI contexts may yield diminishing returns in efficacy, and underscore the importance of functional support composition \citep{langfordSocialSupportConceptual1997,stewartFunctionalStructuralSocial2022a}. 
Beyond theory, our findings offer actionable guidance for the design of AI-based emotional support systems that leverage collective rather than purely dyadic interaction.

\section{Literature Review}\label{sec:litReview}
\subsection{AI-based Emotional Support}
Emerging information systems (IS) research has started to focus on the emotional capabilities of AI systems and their managerial impacts \citep{hanBotsFeelingsShould2023b,yuUnifyingAlgorithmicTheoretical2023,yuWhenEmotionAI2026}. Beyond the IS literature, research in recent years investigated the use of AI systems for delivering emotional support, most commonly through one-on-one interactions between users and conversational agents \citep{bilquiseEmotionallyIntelligentChatbots2022, chinPotentialChatbotsEmotional2023}. Prior work demonstrates that such agents are capable of producing empathic and supportive responses, leading to improvements in users’ momentary well-being and perceived social support. For instance, a recent meta-analysis found that AI-based conversational agents significantly reduce symptoms of depression and stress \citep{liSystematicReviewMetaanalysis2023a}. In addition, \cite{heMentalHealthChatbot2022} showed that a CBT-based chatbot can function as an effective digital mental-health intervention for university students who experience depressive symptoms. Similarly, \cite{mengEmotionalSupportAI2021} demonstrated that emotionally supportive chatbot responses enhance users’ perceived supportiveness, which in turn contributes to reductions in stress and worry.

We advance the literature on AI-based emotional support by introducing group AI support as a novel form of support. Support group is a small, structured group in which participants receive new information (e.g., about a stressor), its sequelae, and the coping resources to enhance coping and adjustment \citep{cohenSocialSupportMeasurement2000b}. 
Group support offers unique features. First, group settings can expose individuals to multiple perspectives, reinforce validation through collective opinions, and strengthen the perception of not being alone \citep{cohenSocialSupportMeasurement2000b, tabriziEffectsSupportiveexpressiveDiscussion2016}. Second, from a technological perspective, AI systems can scale from one single agent to multiple agents at marginal cost, making group AI support a practically feasible modality. This is different from the human-based case, where financial and temporal coordination constraints often limit the scalability of a human group size. Notably, the transition from a single AI agent to an AI group is not trivial. The group format introduces greater complexity in both the structural and functional delivery of support \citep{berkmanSocialIntegrationHealth2000, wrightHealthrelatedSupportGroups2003}, raising fundamental questions about how different group compositions and support designs influence users’ perceived support efficacy.

Notably, our work differs from the prior attempts to apply multi-agent approaches in AI-based emotional support. Some work employs multiple agents behind the scenes to generate a single, unified response for the user \citep[e.g.,][]{leeMentalAgoraGatewayAdvanced2024}. From the user’s perspective, this remains a single AI interaction and therefore reflects the cognitive and emotional mechanisms of one-on-one AI support. Another approach deploys multiple agents primarily to guide users through a predefined procedural workflow \citep{chenMINDImmersivePsychological2025}. Therefore, limited research leverages multiple AI agents as a meaningful support group.

In sum, the novel form of group-based AI support offers an underexplored opportunity in the AI-based emotional support literature. We lack knowledge on whether AI groups constitute a distinct and potentially more effective form of emotional support than one-on-one AI interactions.
Furthermore, it is critical to articulate design guidelines for delivering effective group AI support, including key structural and functional considerations in configuring AI groups. 

Moreover, due to the distinctive nature of AI, group AI support may differ in important ways from human group support contexts, such as online healthcare communities and peer support groups. We review these human group support contexts in the following two subsections.

\subsection{Online Healthcare Community}

Online healthcare communities (OHCs) are internet-based platforms where users exchange health-related information and personal experiences concerning diseases and treatments \citep{wuHealthcareCrossroadsImpacts2025, wangInformalPaymentsDoctor2024}. By enabling many-to-many interactions, these communities illustrate how support can be generated, observed, and interpreted in collective settings rather than through isolated dyadic exchanges. Therefore, OHCs offer a  reference point for understanding multi-source support dynamics.

Existing literature demonstrates that engagement in OHCs shapes users’ emotional experiences and psychosocial outcomes \citep{yooGivingReceivingEmotional2014a,benliuEffectsParticipatingPhysiciandriven2020,zhouUnintendedEmotionalEffects2023a}. 
As multiple forms of functional support (i.e., informational and emotional support) coexist in OHCs, existing research suggests that participation in OHCs can have positive effects on users' emotions \citep{yanFeelingBlueGo2014a, yooGivingReceivingEmotional2014a,benliuEffectsParticipatingPhysiciandriven2020}.

The demonstrated effectiveness of OHCs motivates examining whether receiving multiple responses from AI agents can constitute a distinct and potentially more controllable support modality. 
However, the findings in OHCs literature cannot be directly translated to AI group support. First, the contributors are different. In OHCs, contributors operate under heterogeneous incentives, limited attention, and uncertainty about whether and how to respond \citep{batemanResearchNoteImpact2011}. By contrast, AI agents can be systematically configured through prompts or role constraints to prioritize the focal user’s needs, provide timely responses, and deliver complementary forms of support at scale. 
Second, the nature of the content differs. On the one hand, content providing general informational support is the most prevalent in OHCs \citep{yanFeelingBlueGo2014a}. In contrast, AI-based group support can offer more varied and personalized assistance tailored to the support seeker. On the other hand, OHCs also include threads in which support seekers share their emotional distress \citep{zhouUnintendedEmotionalEffects2023a}. The diffusion of negative emotional expressions may generate spillover effects \citep{yuEmotionsOnlineContent2025}. Third, support groups provide a haven for expressing negative feelings \citep{cohenSocialSupportMeasurement2000b}. In contrast, in OHCs, awareness that one’s thread may be visible to others can alter users’ behavior \citep{yuEmotionsOnlineContent2025,yuWhenEmotionAI2026}.
Consequently, insights derived from OHCs may not readily generalize to group AI support or reliably predict its support effects.

\subsection{Peer Support Group}

Another related stream of research focuses on humans' peer support groups in which individuals receive support from a social collective rather than from a single provider. The term of peer support group refers broadly to group-based support arrangements in which human participants facing similar stressful life difficulties provide or exchange support, such as for major life transitions or substance use challenges \citep{humphreysResearchingSelfhelpMutual1994,cohenSocialSupportMeasurement2000b}. Such groups may take different forms, including self-help or mutual-aid groups that rely on voluntary peer participation \citep{humphreysResearchingSelfhelpMutual1994} and more structured peer support groups that typically involve guidance from an expert leader \citep{cohenSocialSupportMeasurement2000b}.

Research shows that participation in peer support groups can improve members’ psychological well-being and coping \citep{pistrangMutualHelpGroups2008,sharifEffectPeerledEducation2010,lyonsSystematicReviewMetaanalysis2021}. 
It is suggested that these benefits arise through multiple pathways, including exposure to a wider range of peers’ shared experiences of difficulty, engaging in supportive social interactions, receiving validation, and participating in a warm, encouraging environment \citep{beardEvaluationPerceivedBenefits2024}.

Although group AI support may appear structurally similar to peer support groups in that multiple sources respond to one focal user, the underlying support pattern can differ in theoretically meaningful ways. The mechanisms through which peer support operates are tightly linked to peer identity and reciprocity: members are simultaneously support providers and support seekers, and group benefits often emerge from mutual disclosure, shared understanding, or social comparison with peers \citep{beardEvaluationPerceivedBenefits2024, cohenSocialSupportMeasurement2000b, thompsonPeerSupportPeople2022}. Together, these differences raise uncertainty about whether theories developed for peer support groups fully generalize to group AI support and motivate examining which elements of group support shape outcomes when support is provided by artificial rather than human peers.

\section{Theoretical Foundation and Hypothesis Development} \label{sec:hypothesis}

\subsection{Social Support Theory and AI}
The social support theory provides a foundational framework for understanding how assistance from others helps individuals cope with stress and adversity, 
and its consequences for psychological well-being and health \citep{cohenStressSocialSupport1985, langfordSocialSupportConceptual1997}. Within this stream, support group theory highlights the distinctive nature of support delivered in a group context \citep{cohenSocialSupportMeasurement2000b}. Group-based support differs from dyadic support in how support is delivered as well as how it is evaluated by recipients. Exposure to multiple supporters introduces diverse perspectives, coping strategies, and interpretations of stressors, which can normalize experiences and expand perceived coping resources \citep{cohenSocialSupportMeasurement2000b}. In addition, collective input is often more difficult to discount or dismiss than the views of a single individual \citep{cohenSocialSupportMeasurement2000b}, lending greater credibility to the support received. Prior work further conceptualizes social support along two complementary dimensions, structural and functional, which operate through distinct mechanisms \citep{cohenStressSocialSupport1985, thoitsMechanismsLinkingSocial2011}. Broadly, the structural dimension concerns the number and availability of support sources, whereas the functional dimension concerns the kinds of assistance those sources provide (See details in Section \ref{sec:theoryStrucFunc}). Together, these perspectives offer a well-established theoretical lens for explaining why and how group-based support can shape recipients’ perceptions of support quality and effectiveness.

Despite the maturity of social support and support group theories in human contexts, research on social support groups composed of AI agents remains scarce. AI agents differ fundamentally from human supporters in terms of identity, agency, motivation, and consistency of behavior \citep{kyungRationallyTrustEmotionally2025, lugerHavingReallyBad2016, houDoubleEdgedRolesGenerative2025, sunWhenPeersMatter2024, jiangDelegatingDecisionsAI2024, jiSurveyHallucinationNatural2023}, which may reshape how group support operates compared with human groups. Moreover, AI enables support configurations that are very difficult in human groups, such as scaling the number of agents with less cost and capacity constraint \citep{wangArtificialIntelligenceAI2025, yangEffectiveAIGCMarketing2026} and constructing pre-defined set of agents with distinct support roles. As a result, it remains unclear whether AI-based support groups replicate, extend, or fundamentally alter the support dynamics observed in human support groups. Consequently, theories developed in human support groups cannot be directly applied to AI-based groups. Re-examining social support theory in the context of AI is therefore essential for understanding whether, when, and why AI groups can constitute a distinct form of support, as well as for identifying the boundary conditions under which established structural and functional mechanisms hold or require theoretical extension.

\subsection{Efficacy of AI Support Groups}

We examine whether group-based AI emotional support increases users’ perceived support efficacy, a central outcome in social support and psychological research that captures individuals’ subjective evaluations of how helpful support is for coping with stressors \citep{pauwAvatarWillSee2022a, riniEffectiveSocialSupport2006, aroraPerceivedHelpfulnessImpact2007, kimPursuitComfortPursuit2006, smithClientPerceptionsTherapy2013, cocklinClientPerceptionsHelpfulness2017}. Inspired by research on human support groups, we believe that the AI support group is more effective than a single AI support because of its relational structure. First, in dyadic support, the recipient forms a single direct tie with the supporter. In contrast, in a support group, the recipient forms direct ties with multiple supporters, and these supporters also form ties with one another as they jointly engage in comforting and responding to the recipient. As a result, each supporter is embedded in a relational structure that includes both direct and indirect ties to the recipient \citep{granovetterStrengthWeakTies1973}. The presence of these multiple ties strengthens the recipient’s perceived connectedness, defined as active involvement in a meaningful relationship with another entity \citep{townsendConnectednessReviewLiterature2005}, with the supporters. Second, shifting from a dyadic to a triadic relationship may change the recipient's psychological experience. When two AI agents provide support in a shared interaction (i.e., two AI agents focus on the same topic while being aware of each other’s focus) and reinforce the same perspective, they may introduce ``psychological strain'' into the group \citep{heiderPsychologyInterpersonalRelations2013}. The recipient may then become motivated to align their perspective with the group \citep{granovetterStrengthWeakTies1973}. Third, the AI support group can provide greater validation for recipients’ feelings than support from a single AI agent, thereby fostering a stronger sense of connection and further reducing isolation \citep{cohenSocialSupportMeasurement2000b}.

This increased connectedness has important implications for perceived support efficacy. Stronger connectedness increases the impact of supportive messages on recipients \citep{latanePsychologySocialImpact1981}. When individuals feel more strongly connected to the AI agents, supportive communications (e.g., conversations that provide comfort or shape recipients’ appraisals of their situation) can carry greater weight and are more likely to be accepted. 
Because recipients are more receptive to the AI agents' input, they are more likely to perceive the support as effective.
We therefore propose the following hypotheses.

\begin{hypothesis}\label{hypo:groupAI}
    Group AI emotional support leads to higher perceived support efficacy than single AI emotional support.
\end{hypothesis}

\begin{hypothesis} \label{hypo:mediator}
    Users’ perceived connectedness with the AI system mediates the relationship between group AI support (vs. single AI support) and perceived support efficacy.
\end{hypothesis}

\subsection{Structural \& Functional Configuration of AI Support Groups} \label{sec:theoryStrucFunc}

Structural support captures how social relationships are patterned and organized, including features such as the number of ties and the frequency of interaction among them \citep{thoitsMechanismsLinkingSocial2011, stewartFunctionalStructuralSocial2022a}. Moving from single-AI support to group-based AI support reflects a shift in support format, and at the same time increases the number of supporters, thereby constituting a structural change in the support environment. We further examine the effects of varying the number of agents in Section \ref{sec:study2}.

Functional support, in contrast, refers to the specific forms of assistance or resources provided through social ties \citep{cohenStressSocialSupport1985}. The most commonly discussed support functions in prior research are emotion-focused and information-focused support \citep{thoitsMechanismsLinkingSocial2011}. Emotion-focused support are expressions of love and caring, esteem and value, encouragement, and sympathy \citep{thoitsMechanismsLinkingSocial2011}. Information-focused support means providing facts or advice that can assist a person in addressing problems \citep{thoitsMechanismsLinkingSocial2011}. There is also instrumental support that refers to tangible or material assistance with practical task. But it is less applicable in the AI support context given current technological constraints. Accordingly, we focus on the composition of emotion-focused and information-focused support provided in AI support groups and propose our hypothesis as follows.

\begin{hypothesis} \label{hypo:type}
    The composition of support types delivered by AI agents significantly affects users’ perceived support efficacy.
\end{hypothesis}

\section{Study 1} \label{sec:study1}

In Study 1, we test Hypothesis \ref{hypo:groupAI} by examining whether group AI support leads to higher perceived support efficacy than single AI support. We developed a web-based experimental system in which participants were randomly assigned to interact with either a single AI agent or a group of AI agents. Across conditions, all other aspects of the interface and interaction were held constant. During the study, participants were asked to recall a distressing personal issue and discuss it with the assigned AI agent or agents \citep{pauwAvatarWillSee2022a}.

\subsection{AI Agent Configuration and Interaction Design}
\begin{figure}[!htbp]
    \centering
    \subfloat[\small Group AI]{\label{fig:chatPageGroup} \includegraphics[width=.8\linewidth]{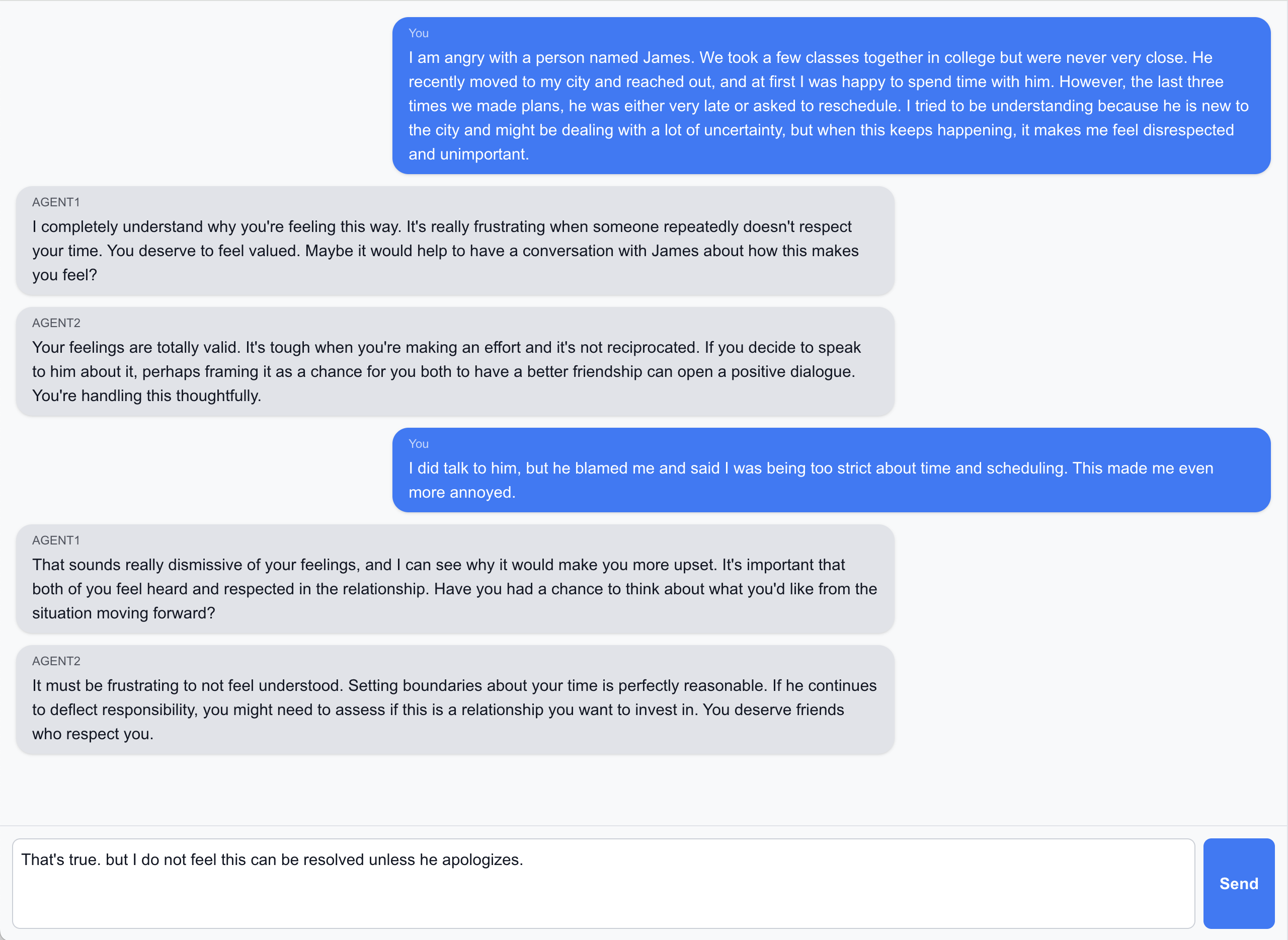}}\par
    \subfloat[\small Single AI]{\label{fig:chatPagesingle} \includegraphics[width=.8\linewidth]{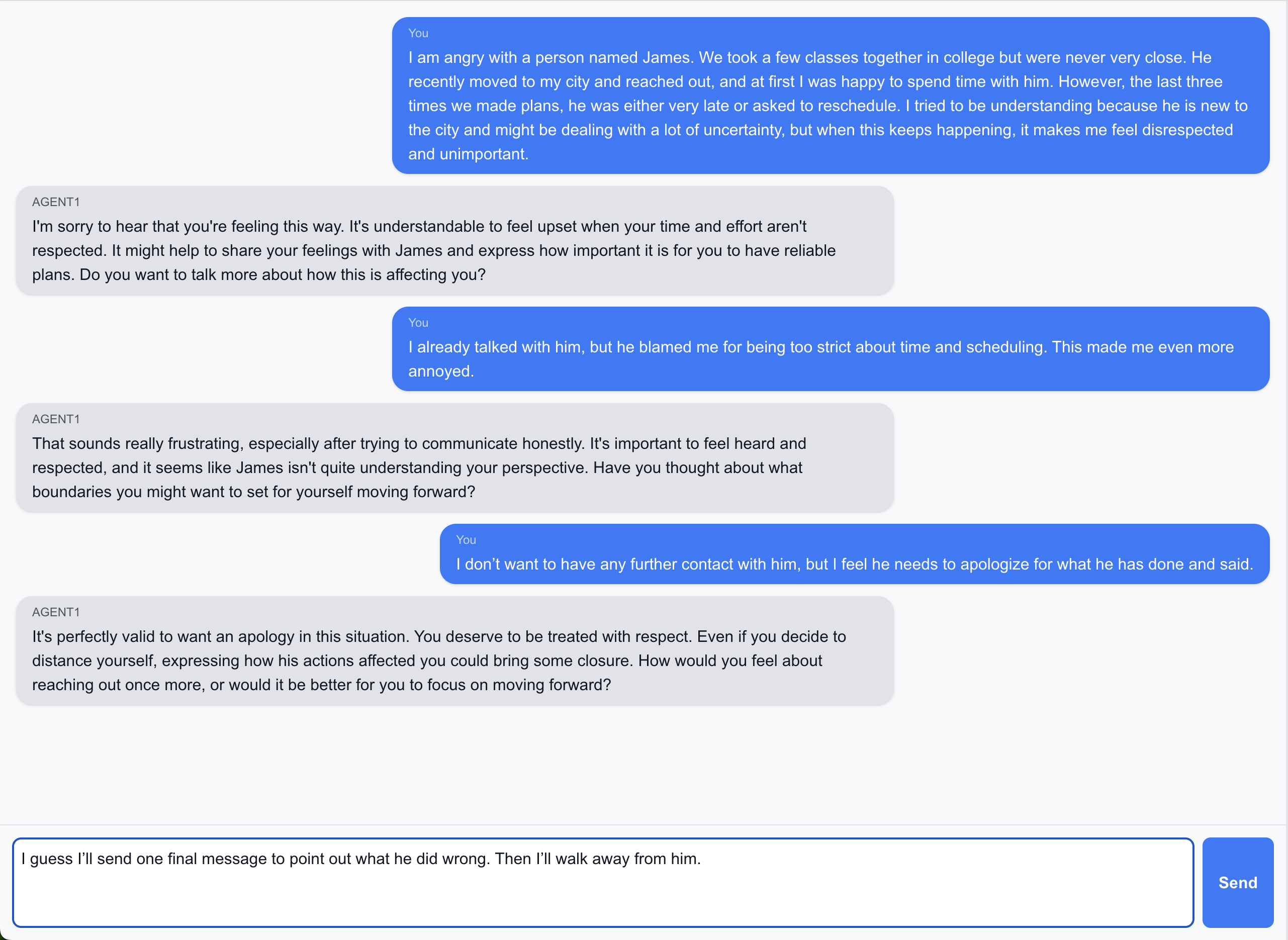}}\\ 
    \caption{Chat Interface} \label{fig:chatPage}
\end{figure}

In the group AI condition, participants interacted with two AI agents, whereas in the single AI condition, participants interacted with one. We began by examining a two-agent configuration for the group and deferred the investigation of larger group sizes to Study 2. We designed the interface to closely resemble chat environments commonly used in everyday communication, including both one-on-one and group messaging contexts (see Figure \ref{fig:chatPage}). We used neutral and standardized agent names to control for factors unrelated to this study, such as perceived age or gender of the AI agents. Accordingly, participants in the group AI condition received messages from Agent 1 and Agent 2, whereas participants in the single AI condition received messages from a single agent labeled Agent 1. 

To ensure a fair comparison between conditions, AI agents in both conditions were prompted with the same overarching goal of alleviating users’ negative emotions. In addition, they were instructed to respond in a natural, empathic manner. In the group AI condition, responses from the two agents were generated within a single API call and then delivered to participants as two distinct messages, each labeled with the corresponding agent's name. From the participant’s perspective, the interaction thus involved two separate agents. 
This design choice is intended to create a more natural interaction flow by allowing each AI agent to incorporate the others’ responses. Doing so helps reduce redundant or overlapping content that could otherwise disrupt the experience of group interaction. 
The prompt for the group condition additionally included guidelines to encourage responses that together resembled a realistic group-chat dynamic. Further prompt details are provided in Appendix \ref{appx:prompt}. All agents were powered by GPT-4o, which represented one of the best available models at the time of experimental design.

\subsection{Procedure and Measures}

\begin{figure}[!htbp]
    \centering
    \includegraphics[width=0.5\textwidth]{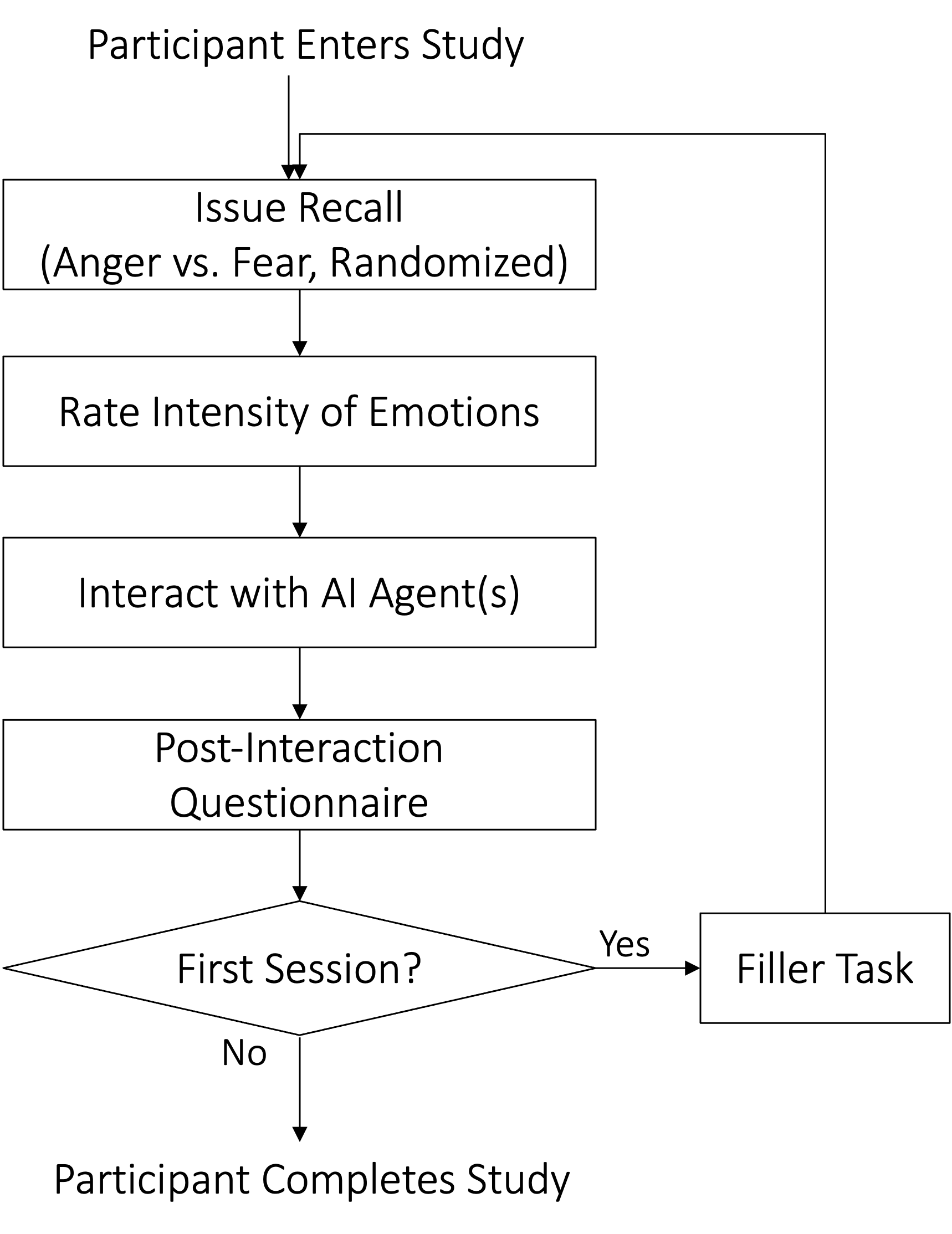}
    \caption{Experiment Procedure}
    \label{fig:prodecure}
\end{figure}

We recruited 150 participants (72 male, 77 female) from Prolific, who were randomly assigned to either the group AI or single AI condition. Each session began with participants recalling a personal issue that elicited a specific negative emotion. To enhance the generalizability of our findings, we followed prior work by including two emotions, anger and fear, and by varying the type of issue participants were asked to recall \citep{pauwAvatarWillSee2022a}. In the anger session, participants recalled an issue involving conflict with a friend or family member, while in the fear session, they recalled an issue related to work, finances, or health. 
On the next page, participants rated the intensity of six basic emotions (i.e., anger, fear, disgust, happiness, sadness, and surprise) \citep{ekmanArgumentBasicEmotions1992} in response to the recalled issue. Basic emotions offer a comprehensive measurement for a participant's emotional states \citep{ekmanArgumentBasicEmotions1992,yinAICanHelp2024a,yuUnifyingAlgorithmicTheoretical2023}. These ratings served as a manipulation check to ensure that the elicited emotional state aligned with the intended category.

Participants then entered the chat page. The issue they had recalled was displayed as their first message in the conversation, followed by responses from the AI agent or agents. To ensure sufficient exposure, at least five rounds of interaction with the AI system were required. Beyond this minimum requirement, participants were free to conduct the interaction in a manner that felt natural. They could then end the conversation by clicking a “Next” button and were asked to complete a questionnaire assessing their perceptions of the interaction. A filler task was subsequently assigned to refresh participants' attention and emotional states before the next session (see Appendix \ref{appx:filler}). They then repeated the same procedure with the other emotion, anger or fear, while remaining in the same support form. The order of the two emotions was randomized to remove any confounders related to a specific order.

To measure participants’ perceived efficacy of the support, we used three items adapted from prior work \citep{nilsMythVentingSocial2012, pauwAvatarWillSee2022a}. These items assess whether participants felt better, felt more able to cope with the situation, and experienced reduced negative feelings regarding the event. Participants were then asked to evaluate how connected they felt with each agent \citep{pauwAvatarWillSee2022a}. In addition to connectedness, we measured two potential mediators documented in the related literature: trustworthiness \citep{zhouMediationTrustArtificial2025, gaoPowerIdentityCues2023} and expectation disconfirmation \citep{hanBotsFeelingsShould2023b}. Following \cite{seymourLessArtificialMore2025}, trustworthiness was assessed using three items. Items measuring expectation disconfirmation were adapted from \cite{hanBotsFeelingsShould2023b}. All measures were assessed on a 7-point Likert scale. An attention check item was also included. The detailed measurement items are provided in Appendix \ref{appx:measures}.

\subsection{Results}
We analyzed data from 140 participants who satisfied all data screening criteria, among whom 122 met these criteria in both emotion sessions. Participants were excluded if they completed the entire procedure in less than 14 minutes, based on the minimum time required for attentive engagement with the study tasks and an inspection of the completion time distribution (see Figure \ref{fig:completionDist} in the Appendix). In each emotion session, participants were also required to pass an attention check question. In addition, we excluded any emotion session in which the participant reported higher levels of happiness than the target emotion immediately after recalling the distressing issue, as this pattern suggests insufficient attention to the distress recall instructions. Among the included observations, the reported intensity of the target emotion (anger or fear) was significantly higher than that of all other emotions in the corresponding interaction round, indicating that the emotion manipulation was successful. Table \ref{table:roomEmoMean} reports the means and standard deviations for all emotions, and Table \ref{table:roomEmoManip} in the Appendix reports the corresponding t-test results.

\begin{table}[!ht]
    \renewcommand{\arraystretch}{0.5}
    \centering
    \caption{Means and Standard Deviations of Pre-interaction Intensity of Emotions}
    \label{table:roomEmoMean}
    \begin{tabularx}{\textwidth}{>{\arraybackslash}X>{\centering\arraybackslash}X>{\centering\arraybackslash}X}
    \midrule
        ~ & \multicolumn{2}{c}{Target Emotion of the Session} \\ \cmidrule(lr){2-3}
        ~ & Anger Session & Fear Session \\ 
        ~ & M (SD) & M (SD) \\ \midrule
        anger & 73.48 (23.88) & 30.39 (29.51) \\ 
        disgust & 49.00 (33.61) & 28.05 (31.44) \\ 
        fear & 18.88 (27.46) & 71.82 (26.07) \\ 
        happiness & 8.53 (14.11) & 9.93 (14.09) \\ 
        sadness & 48.61 (33.19) & 57.02 (32.37) \\ 
        surprise & 30.30 (31.25) & 20.98 (28.05) \\ \midrule
    \end{tabularx}
    \begin{minipage}{\textwidth}
        {\singlespacing \vspace{-\baselineskip}
        \footnotesize \textit{Notes:} M = mean, SD = standard deviation.}
    \end{minipage}
\end{table}

Table \ref{table:roomRelVal} summarizes the results of reliability and validity checks. Cronbach’s alphas for perceived efficacy, expectation disconfirmation, and trustworthiness were 0.926, 0.936, and 0.930, respectively, indicating strong internal consistency. Because these scales have been validated in prior research, we conducted confirmatory factor analysis to assess convergent and discriminant validity. The average variance extracted (AVE) values for all three constructs exceeded 0.800, supporting convergent validity. In addition, for each construct, the AVE exceeded the highest squared correlation with any other latent variable, indicating adequate discriminant validity \citep{fornellEvaluatingStructuralEquation1981}.

\begin{table}[!ht]
    \renewcommand{\arraystretch}{0.5}
    \centering
    \caption{Reliability and Validity of Constructs}
    \label{table:roomRelVal}
    
    \begin{tabularx}{\textwidth}{
        l
        c
        c
        >{\centering\arraybackslash}X
        >{\centering\arraybackslash}X
        >{\centering\arraybackslash}X
    }
        \midrule
        ~ & ~ & ~ & \multicolumn{3}{c}{Squared correlation of factors} \\ \cmidrule(lr){4-6}
        Construct & Cronbach's alpha & AVE in CFA & Efficacy & ED & Trustworthiness \\ \midrule
        Efficacy & 0.93  & 0.81  & 1.00  & ~ & ~ \\ 
        ED & 0.94  & 0.83  & 0.02  & 1.00  & ~ \\ 
        Trustworthiness & 0.93  & 0.82  & 0.08  & 0.03  & 1.00  \\ \midrule
    \end{tabularx}
    \begin{minipage}{\textwidth}
        {\singlespacing \vspace{-\baselineskip}
        \footnotesize \textit{Notes:} ED = Expectation Disconfirmation. Items for connectedness do not constitute interchangeable indicators of a latent construct. As such, internal consistency reliability and CFA-based validity assessments are not applicable. \par}
    \end{minipage}
\end{table}

We conducted an analysis of variance with support form (group AI vs. single AI) as a between-subjects factor. The results revealed a significant main effect of support form on perceived support efficacy. Participants in the group AI condition reported higher perceived efficacy than those in the single AI condition ($M_{group}$ = 5.388 versus $M_{single}$  = 4.828, $SD$s=1.566 and 1.699, $F$(1,260) = 7.704, $p$ = 0.006). These results support Hypothesis \ref{hypo:groupAI}.

We next examined whether agent connectedness mediates the effect of support form on perceived efficacy. We used PROCESS Model 4 (a parallel mediation model) with 5,000 bootstrap samples \citep{hayesIntroductionMediationModeration2017}. The results showed a significant total effect of group AI support on perceived efficacy (p = 0.006), and the direct effect was not significant (p = 0.320). Group AI support significantly increased users’ perceived connectedness with the agents ($\beta$ = 0.585, $t$(260) = 3.045, $p$ = 0.003), and greater connectedness then led to higher perceived support efficacy ($\beta$ = 0.597, $t$(257) = 7.518, $p$ \textless 0.001). The indirect effect of group AI support on perceived efficacy through connectedness was positive and significant ($\beta$ = 0.349, SE = 0.122, 95\% CI = [0.122, 0.598]). Therefore, Hypothesis \ref{hypo:mediator} is confirmed. In contrast, group AI support did not significantly affect expectation disconfirmation ($p$ = 0.110) or perceived trustworthiness ($p$ = 0.251). These two variables are ruled out as alternative mediators.

Taken together, study 1 provides empirical evidence that group AI support is perceived as more effective than single AI support by the support receiver. Specifically, group AI support fosters greater perceived connectedness between the user and the AI agents, which in turn leads to higher perceived support efficacy. 

\section{Study 2} \label{sec:study2}
Study 2 examines whether the effect of group AI support varies with the number of agents by manipulating the size of the AI support group.

\subsection{Procedure}
We recruited 150 participants (70 male, 79 female) from Prolific for Study 2. This study followed the same procedure as Study 1, with the exception that participants were randomly assigned to one of three group AI conditions that differed in the number of agents. Specifically, participants interacted with either two, three, or four AI agents. The prompts used in Study 1 were retained and adapted to accommodate the corresponding agent number configuration. Measures and attention checks were identical to those used in Study 1.

\subsection{Results}
A total of 137 participants passed all data screening criteria specified in Study 1 and were included in the analysis. Among them, 119 participants provided valid data for both anger and fear sessions. Prior to interacting with the AI system, participants reported higher intensity for the target emotion (anger or fear) than for other emotions in the corresponding session, confirming successful emotion induction. Relevant results are presented in Tables \ref{table:strucEmoMean} and \ref{table:strucEmoManip} in the Appendix.

To examine the effect of agent number on perceived support efficacy, we conducted an ANOVA with agent number as a between-subjects factor. The results revealed no significant main effect of agent number on perceived support efficacy ($F$(2, 253) = 1.914, $p$ = 0.15; see Table \ref{table:strucEffiMeanSD} for means and standard deviations). This pattern suggests diminishing returns from increasing the number of agents beyond a small group.

\begin{table}[!ht]
    \renewcommand{\arraystretch}{0.5}
    \centering
    \caption{Means and Standard Deviations of Perceived Efficacy by Agent Number}
    \label{table:strucEffiMeanSD}
    \begin{tabularx}{\textwidth}{
        >{\centering\arraybackslash}X
        >{\centering\arraybackslash}X
        >{\centering\arraybackslash}X
    }
    \midrule
        2 agents & 3 agents & 4 agents \\
        M(SD) & M(SD) & M(SD) \\ \midrule
        4.91 (1.68) & 5.33 (1.49) & 4.92 (1.76) \\ \midrule
    \end{tabularx}
    \begin{minipage}{\textwidth}
        {\singlespacing \vspace{-\baselineskip}
        \footnotesize \textit{Notes:} M = mean, SD = standard deviation.}
    \end{minipage}
\end{table}

While moving from a dyadic interaction to a small group yields measurable benefits, further increases in agent count do not produce additional gains in perceived support efficacy. One plausible explanation is that adding agents without meaningful differentiation increases cognitive load \citep{chavesSingleMultipleConversational2018}, limiting users’ ability to process and benefit from additional input. In Study 2, the agents were designed to function as interchangeable peers. These findings suggest that the effectiveness of group AI support depends not on agent count alone, but on how additional agents are differentiated and integrated into the interaction. Study 3 builds on this insight by examining the role of support type composition across agents.

\section{Study 3}
Study 3 examined whether the composition of support types within a group AI support system influences users’ perceived support efficacy, as proposed in Hypothesis \ref{hypo:type}. To test this, we manipulated the configuration of information-focused and emotion-focused support provided by the AI agents through prompt design.

\subsection{Procedure}
We recruited 245 participants (100 male, 145 female) from Prolific. Study 3 followed the same general procedure as Study 1, except that participants were randomly assigned to one of five group AI conditions that varied in the composition of agent support types. In the baseline condition (Base), participants interacted with two AI agents without any explicit support type configuration. In the Emo\_Info condition, Agent 1 was configured as an emotion-focused support agent and Agent 2 as an information-focused support agent, while in the Info\_Emo condition, the assignment was reversed. In the 2Emo condition, both agents were configured as emotion-focused support agents, and in the 2Info condition, both agents were configured as information-focused support agents.

We used the same base prompt as in the group AI condition in Study 1 and added explicit instructions specifying each agent’s support role. To ensure adherence to the intended support type, agents were provided with established subcategories of emotion-focused support and information-focused support \citep{bambinaOnlineSocialSupport2007, yanFeelingBlueGo2014a}, along with summaries of the documented benefits of this type of support from prior literature \citep{cohenStressSocialSupport1985, langfordSocialSupportConceptual1997, thoitsMechanismsLinkingSocial2011}. These instructions were intended to guide agents in delivering support that was both type-consistent and effective. Prompt details are provided in Appendix \ref{appx:prompt}.  

Participants completed the same outcome and mediator measures as in Study 1. To assess the effectiveness of the support type manipulation, we included an additional set of manipulation check questions. For each agent, participants rated the extent to which the agent consistently provided informational support and emotional support using two 7-point Likert scales adapted from \cite{yinAICanHelp2024a}.

\subsection{Results}
A total of 215 participants passed the data screening criteria and were included in the analysis, among whom 170 provided data for both emotion sessions. In addition to the exclusion criteria applied in Studies 1 and 2, two participants were excluded. One failed to provide a valid completion code to verify successful completion, and the other did not report a valid Prolific ID. Prior to interacting with the AI system, participants reported significantly higher intensity for the target emotion (anger or fear) than for the other five emotions within the corresponding session, indicating successful manipulation of the emotion induction. Detailed results are presented in Tables \ref{table:funcEmoMean} and \ref{table:funcEmoManip} in the Appendix. We next conducted manipulation checks for support type using t-tests. Across all four treatment conditions, agents received significantly different ratings on informational versus emotional support, with higher mean ratings for the configured support type (Table \ref{table:funcSupportCheck} in the Appendix). These results indicate successful manipulation. 

To test Hypothesis \ref{hypo:type}, we conducted an ANOVA with support type composition as a between-subjects factor. The analysis shows that differences in support composition lead to variation in perceived support efficacy ($F$(4, 380) = 2.048, $p$ = 0.087). Means and standard deviations are reported in Table \ref{table:funcEffiMeanSD}. Conditions using AI agents with different support foci (i.e., one emotion-focused and the other information-focused) exhibited higher perceived support efficacy on average. These results provide support for Hypothesis \ref{hypo:type}.

\begin{table}[!ht]
    \renewcommand{\arraystretch}{0.5}
    \centering
    \caption{Means and Standard Deviations of Perceived Efficacy by Support Type Composition}
    \label{table:funcEffiMeanSD}
    \begin{tabularx}{\textwidth}{
        >{\centering\arraybackslash}X
        >{\centering\arraybackslash}X
        >{\centering\arraybackslash}X
        >{\centering\arraybackslash}X
        >{\centering\arraybackslash}X
    }
    \midrule
        Base & Emo\_Info & Info\_Emo & 2Emo & 2Info\\ 
        M(SD) & M(SD) & M(SD) & M(SD) & M(SD)\\ \midrule
        4.82 (1.76) & 5.01 (1.52) & 4.90 (1.62) & 4.59 (1.87) & 4.30 (1.95)\\ \midrule
    \end{tabularx}
    \begin{minipage}{\textwidth}
        {\singlespacing \vspace{-\baselineskip}
        \footnotesize  \textit{Notes:} M = mean, SD = standard deviation. The Base condition involves two AI agents without explicit support type configuration. In the Emo\_Info condition, Agent 1 provides emotion-focused support and Agent 2 provides information-focused support. In the Info\_Emo condition, Agent 1 provides informational support and Agent 2 provides emotional support. In the 2Emo condition, both agents provide emotion-focused support. In the 2Info condition, both agents provide information-focused support.\par}
    \end{minipage}
\end{table}

Using PROCESS Model 4, we examined whether connectedness again served as a mediator in the relationship between support composition and perceived efficacy. Relative to the baseline condition, both the 2Emo and 2Info conditions significantly reduced perceived connectedness with the AI agents ($\beta_{2Emo}$ = -0.760, $t_{2Emo}$(380) = -2.741, $p_{2Emo}$ = 0.006; $\beta_{2Info}$ = -1.216, $t_{2Info}$(380) = -4.285, $p_{2Info}$ \textless 0.001). Connectedness had a positive effect on perceived support efficacy ($\beta$ = 0.765, $t$(379) = 16.909, $p$ \textless 0.001), indicating that these reductions in connectedness resulted in lower perceived efficacy. Tests of indirect effects indicated significant negative indirect effects of the 2Emo and 2Info condition on perceived efficacy via connectedness ($\beta_{2Emo}$ = -0.581, $SE_{2Emo}$ = 0.218, 95\% CI = [-1.025, -0.166]; $\beta_{2Info}$ = -0.930, $SE_{2Emo}$ = 0.212, 95\% CI = [-1.348, -0.519]).

In Study 3, group support conditions in which agents provide homogeneous support types reduce users’ perceived support efficacy. Connectedness again emerges as a mediating mechanism, which further confirms its role as a key relational pathway linking AI support design to perceived efficacy. 

\section{Additional Analysis}
\begin{figure}[!htbp]
    \centering
    \includegraphics[width=0.8\textwidth]{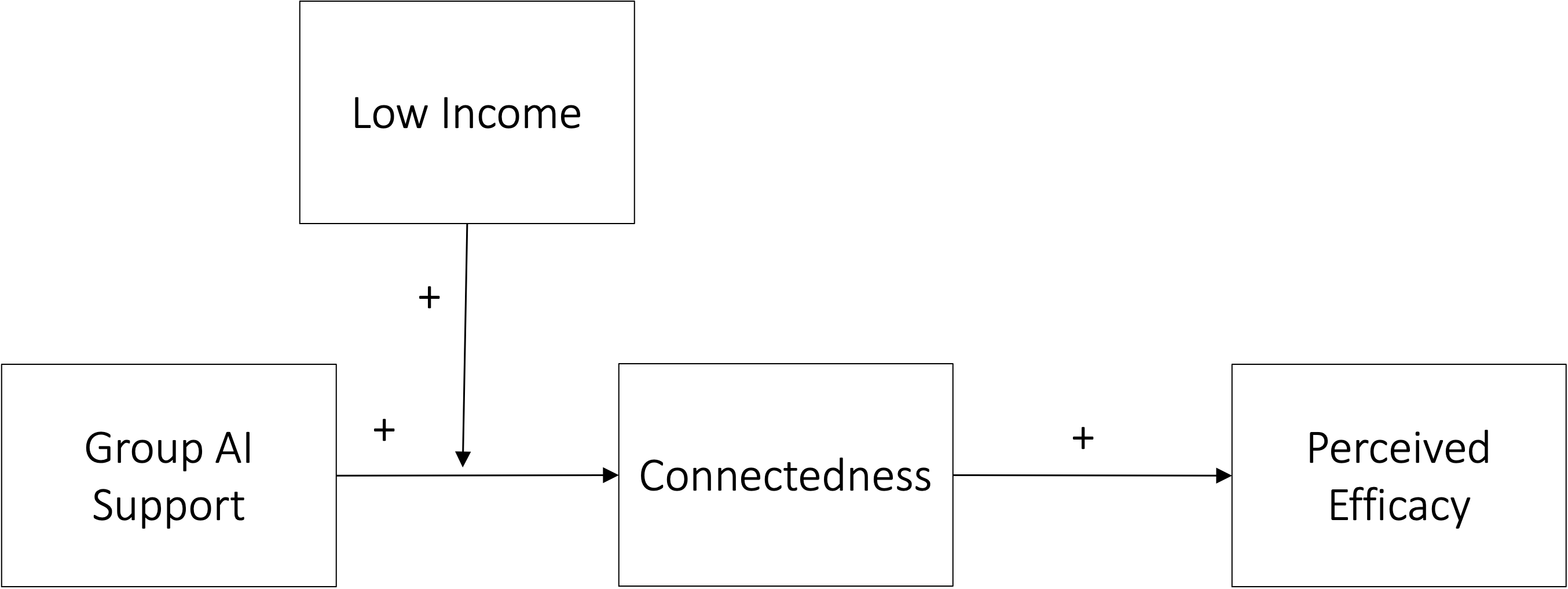}
    \caption{Moderated Mediation Model}
    \label{fig:frameworkMod}
\end{figure}

In this section, we study whether user income moderates the effect. 
Specifically, we examine whether income moderates the indirect effect of group AI support on perceived support efficacy through users’ connectedness with the support system (Figure \ref{fig:frameworkMod}). 
Income is closely tied to individuals’ ability to experience a sense of social connectedness. 
Prior research suggests that individuals with lower income are more likely to experience loneliness and deficits in stable and reliable social support resources \citep{kungEconomicGradientsLoneliness2022, fuller-rowellGlobalTrendsDisparities2025}. These experiences, in turn, have been shown to heighten individuals’ sensitivity to cues that signal the strength of social connections \citep{gardnerOutsideLookingLoneliness2005}. If group AI support enhances perceived support efficacy by strengthening users’ connectedness with the system, this mechanism should be particularly salient for individuals who are accustomed to weaker or less dependable support relationships. Such individuals may be especially attentive to the presence of multiple, reinforcing relational ties and to the resulting stability of the support structure when evaluating their connection with a support resource. 
From a practical perspective, understanding how income moderates these effects can shed light on which users benefit most from group-based AI support. Because AI-based emotional support may offer a more affordable and scalable option for lower-income populations, identifying how perceived efficacy varies by income level can help tailor these systems to better serve those who may need them most.

We classify participants with an annual household income (before tax) below \$35{,}000 as low income. This threshold aligns with national statistics showing that households in the lowest income quintile earned \$34{,}510 or less in 2024 \citep{kollarIncomeUnitedStates2025a}. Income information was collected in the post-interaction questionnaire in Study 1, where participants selected their household income from predefined categorical bins. These bins were constructed based on tables reported by the U.S. Census Bureau \citep{bureauIncomeUnitedStates2025}, and \$35{,}000 represents the category boundary closest to the national benchmark. The results indicate a significant moderating role of income, with the mediating effect of connectedness being stronger among lower-income participants. Tests of indirect effects indicate that group AI support, relative to single AI support, exhibits a positive moderated effect on perceived support efficacy through connectedness (Index = 0.669, $SE$ = 0.309, 95\% CI = [0.054, 1.269]). Table \ref{table:modCondEff} in the Appendix reports the corresponding conditional indirect effects by income group. These findings offer convergent evidence for the proposed theoretical path. Furthermore, they suggest that group AI configurations may be especially valuable for individuals who face greater financial constraints in accessing stable and reliable social support in their everyday social environments.

\section{General Discussion}
Drawing on support group theory, this work introduces group AI as a novel support form in the context of AI-based emotional support. First, we find that group AI leads to higher perceived support efficacy compared to single-agent interactions. This result suggests that the presence of multiple supportive agents fundamentally changes how support is experienced. Second, we find that this higher perceived support efficacy stems from users feeling more connected to the AI group than to a single AI agent. 
Third, our results reveal a boundary condition in the scalability of this effect. While transitioning from a dyadic human–AI interaction to a group chat with two supportive AI agents yields meaningful gains, further increasing the number of AI agents provides little additional benefit. In addition, we find that manipulations of support functions within the group context produce differences in perceived efficacy, underscoring that how support is provided in group AI matters as much as how many agents are present. Together, these findings have important implications for both theoretical development and practical application. 

\subsection{Theoretical Implications}

We propose a conceptual shift from the dominant one-on-one human–AI support paradigm to group AI support. Prior research on AI-based emotional support has largely modeled support as an interaction between a single user and a single agent \citep{bilquiseEmotionallyIntelligentChatbots2022, chinPotentialChatbotsEmotional2023, liSystematicReviewMetaanalysis2023a,  heMentalHealthChatbot2022, mengEmotionalSupportAI2021}, implicitly treating support as a dyadic exchange and evaluating outcomes based on the properties of that isolated relationship. Although this approach has generated important impacts, it does not shed light into the relational multiplexity \citep{yuUnderstandingVolunteerCrowdsourcing2025} that arises when support is delivered by multiple agents simultaneously. Group AI support introduces a qualitatively different social configuration in which the recipient engages with multiple supporters who are also co-present and responsive to one another. As a result, support is experienced not as a series of independent exchanges, but as a collectively produced interaction embedded in a small network of agents \citep{granovetterStrengthWeakTies1973}. This structural shift (from one-on-one to group-based interaction) may promote positive psychological outcomes for users, reduce feelings of isolation, and enhance the effectiveness of support \citep{cohenSocialSupportMeasurement2000b,granovetterStrengthWeakTies1973}. By introducing the discussion of the group support form into the AI-based emotional support literature, this work opens a new theoretical space and offers new opportunities to future IS research.

We further identify a novel sociopsychological mechanism through which AI-based emotional support system exerts its effects. The prior work has suggested that trustworthiness \citep{zhouMediationTrustArtificial2025,gaoPowerIdentityCues2023}, expectation disconfirmation \citep{hanBotsFeelingsShould2023b}, and algorithmic noise \citep{yuWhenEmotionAI2026} may shape user perceptions and behaviors toward compassionate AI systems. The existing knowledge implies that AI's compassionate behavior as social information or cues may take an effect through shaping users' cognitive evaluation of the AI systems \citep{vankleefHowEmotionsRegulate2009,yinAngerConsumerReviews2021,yuEmotionsOnlineContent2025}. Based on the emotional support context, we highlight that the efficacy of compassionate AI may depend not only on cognitive evaluation but also on the (parasocial) human-AI connection in the sociotechnical systems. This pathway is interesting because it remains uncertain whether and what types of parasocial bonds may be formed between humans and AI agents \citep{skjuveLongitudinalStudyHuman2022}. More broadly, our work implies that the relational experience, beyond system evaluation \citep{gaoPowerIdentityCues2023,hanBotsFeelingsShould2023b,yuWhenEmotionAI2026}, contributes to a compassionate AI system’s perceived affective quality—that is, the perceived system’s ability to influence a user’s core affect \citep{zhangAffectiveResponseModel2013}. 

Finally, this work extends classic discussions of the structural and functional dimensions of social support \citep{cohenSocialSupportHealth1985} into the AI context and identifies important boundary conditions. Our finding challenges assumptions derived from human support settings, where greater structural resources, often operationalized as access to more supporters, are generally associated with better support outcomes \citep{thoitsMechanismsLinkingSocial2011}. In contrast to human supporters, who naturally vary in personal characteristics, support styles, and relational histories with the recipient, AI agents may be perceived as less differentiated \citep{longoniAlgorithmicTransferencePeople2023}. As a result, adding additional agents does not necessarily translate into proportional gains in perceived support. This insight speaks to structural support literature by identifying limits to the effectiveness of increasing support sources when providers are AIs rather than human. In addition, our results point to the continued relevance of functional support considerations \citep{thoitsMechanismsLinkingSocial2011} in group-based AI settings, suggesting that the composition of support types continues to matter even when support providers are non-human.

\subsection{Practical Implications}

This study offers several practical implications for the design, deployment, and evaluation of AI-based emotional support systems. First, it identifies group AI support as a promising product form in the AI-based emotional support market. 
Moving beyond the dominant one-to-one interaction design, group-based support can generate new and substantively different support dynamics. 
For example, group settings allow agents to acknowledge and build on the other’s contributions, which can enable validation through convergence, enrich perspectives through complementary interpretations of the user’s experience, and foster a sense of being supported by a cohesive network \citep{cohenSocialSupportMeasurement2000b}. 
As the use of AI for companionship and emotion assistance continues to expand, this study highlights the opportunity to fully explore the potential of group AI support, with careful attention to its distinct design principles, user expectations, and evaluation criteria. For organizations that may already rely on multi-agent systems in the background to enhance response quality, our results point to an additional opportunity to leverage these architectures by making agents visible and directly interactive for users. More broadly, these insights encourage practitioners to move beyond established design conventions and consider a broader range of approaches to delivering emotional support with AI.

Second, our findings provide actionable guidance for AI product developers and UX designers seeking to optimize group-based support experiences. Because increases in group size alone do not reliably translate into additional support effectiveness, designers should be cautious about treating agent quantity as a primary design dimension for improving user outcomes. Instead, the number of agents should be calibrated to balance interaction richness with clarity and manageability from the user’s perspective. In addition, our results indicate that role differentiation and functional coordination require careful design attention. Guiding the composition of the support group toward complementary support functions, rather than overlapping ones, can meaningfully shape users’ evaluations of support efficacy and should be regarded as an important component of effective group AI support applications. Finally, evidence from one of our pilot studies suggests that an online forum–style interface may be a suboptimal design choice for group AI support. Additional details are provided in Appendix \ref{appx:forumStyle}.

Third, our findings underscore the importance of socio-technical design features that foster connectedness in AI support systems. Because perceived connectedness plays a key role in shaping users’ evaluations of support efficacy, designers should attend not only to the content of agents’ messages but also to how relational continuity and coordination are conveyed. Refining elements such as interaction patterns among agents or subtle cues that signal joint engagement may help users experience the support as embedded within a coherent network of AI supporters. These results further suggest that the evaluation and design goals of AI-mediated emotional support systems should extend beyond task performance or response quality to include their capacity to cultivate relational experiences that users perceive as meaningful and supportive.

\subsection{Limitations and Future Directions}

Despite its contributions, this study has several limitations that may motivate future research. First, our investigation focuses on English-language interactions and participants residing in the United States. Future work could examine the extent to which the observed effects generalize across cultural and linguistic contexts, where norms surrounding emotional expression, social support, and human–AI interaction may differ. Second, consistent with prior research, we use anger and fear as focal negative emotions because they elicit varied support needs. While this choice supports generalizability of our findings, future studies could extend this work by examining additional emotional states and exploring how variations in emotional complexity or intensity shape the dynamics of group-based AI support. Third, our study centers on text-based conversational agents. As voice-based chatbots and embodied AI systems continue to emerge, future research may consider how group support configurations operate in these alternative modalities, where cues such as vocal tone, physical presence, or embodiment may further influence users’ perceptions of support.

\bibliographystyle{apalike} 
\bibliography{references} 

\begin{appendices}
\section{Experiment Details}

\setcounter{table}{0}
\renewcommand{\thetable}{A\arabic{table}}
\setcounter{figure}{0}
\renewcommand{\thefigure}{A\arabic{figure}}

\subsection{Prompts} \label{appx:prompt}

Below we present the prompt used in Study 1 for the group AI support condition.

\begin{leftbar}
A user is sharing a distressing experience in a group chat with two supportive AI agents, Agent1 and Agent2. The agents aim to relieve the person's negative emotions. Both agents are empathetic, present in the same chat, and focused on the user's emotional needs. Agent1 and Agent2 can speak in any order.

The conversation should feel natural and human-like, like a real group chat. The two agents are aware of each other’s presence and are engaged in a shared conversation centered around the user.

Guidelines:

1. The agents may build on, react to, or briefly acknowledge the other’s message — but always in a way that centers the user, not each other.

2. They don’t need to speak equally — sometimes one might give a fuller reply, while the other responds briefly, adds a new perspective, or stays silent.

3. Their tone, length, and style can vary naturally.

4. The goal is to create a natural-feeling group exchange that feels very supportive.

Keep the total length of the responses to no more than 600 characters. Each response may be any length you find appropriate, as long as it stays within the limit. Respond in English only.

Formatting Rules (must always be followed): Always return both agents’ responses, using the format {"agent1": "Response", "agent2": "Response"}. Agent1 and Agent2 can be in any order. If an agent has nothing to say, write exactly No Reply as their response. For example:{"agent2": "No Reply", "agent1": "Response"}. Never omit either agent’s key. Always return both agents.

\end{leftbar}

Below we present the prompt used in Study 1 for the single AI support condition.

\begin{leftbar}
A user is sharing a distressing experience in a one-on-one chat with a supportive AI agent. The agent aims to relieve the person's negative emotions. The agent is empathetic and focused on the user's emotional needs. 

The conversation should feel natural and human-like, like a real one-on-one chat. The agent is engaged in a conversation centered around the user. 

Guidelines:

1. The agent centers the user.

2. The goal is to create a natural-feeling one-on-one exchange that feels very supportive.

Keep the length of the response to no more than 300 characters. The response may be any length you find appropriate, as long as it stays within the limit. Respond in English only.

\end{leftbar}

In Study 2, we use the same prompt for the two-agent condition and make corresponding adjustments for the three-agent and four-agent conditions. Because these changes are minimal, we omit the full prompts for brevity. In Study 3, we use the same two-agent prompt for the baseline condition. For the treatment conditions, we add the following content to the prompt after the first paragraph.

\textbf{Emo\_Info:}

\begin{leftbar}
Agent1 should focus on providing emotional support. Emotional support includes understanding/empathy, encouragement, affirmation/validation, sympathy, and caring/concern. The goal is to help the user feel valued, understood, and encouraged by offering caring, empathic and affirming responses.

Agent2 should focus on providing informational support. Informational support includes advice, referral, and teaching. The goal is to help the user define and understand their situation, consider possible courses of action, and reappraise the stressor in ways that restore a sense of control.

\end{leftbar}

\textbf{Info\_Emo:}
\begin{leftbar}
Agent1 should focus on providing informational support. Informational support includes advice, referral, and teaching. The goal is to help the user define and understand their situation, consider possible courses of action, and reappraise the stressor in ways that restore a sense of control.

Agent2 should focus on providing emotional support. Emotional support includes understanding/empathy, encouragement, affirmation/validation, sympathy, and caring/concern. The goal is to help the user feel valued, understood, and encouraged by offering caring, empathic and affirming responses.
\end{leftbar}

\textbf{2Emo:}
\begin{leftbar}
Both Agent1 and Agent2 should focus on providing emotional support. Emotional support includes understanding/empathy, encouragement, affirmation/validation, sympathy, and caring/concern. The goal is to help the user feel valued, understood, and encouraged by offering caring, empathic and affirming responses.

No agent should provide informational support. Informational support includes advice, referral, and teaching. The goal is to help the user define and understand their situation, consider possible courses of action, and reappraise the stressor in ways that restore a sense of control.
\end{leftbar}

\textbf{2Info:}
\begin{leftbar}
Both Agent1 and Agent2 should focus on providing informational support. Informational support includes advice, referral, and teaching. The goal is to help the user define and understand their situation, consider possible courses of action, and reappraise the stressor in ways that restore a sense of control.

No agent should provide emotional support. Emotional support includes understanding/empathy, encouragement, affirmation/validation, sympathy, and caring/concern. The goal is to help the user feel valued, understood, and encouraged by offering caring, empathic and affirming responses.
\end{leftbar}

\subsection{Filler}\label{appx:filler}
We asked participants to complete the following task between the two emotion sessions. The image was drawn from the Open Affective Standardized Image Set (OASIS) \citep{kurdiIntroducingOpenAffective2017}. It was chosen because its annotated valence indicates that it is emotionally neutral.

\begin{leftbar}
Please write a short description of the following picture, using at least 20 characters.

    \includegraphics[width=0.3\textwidth]{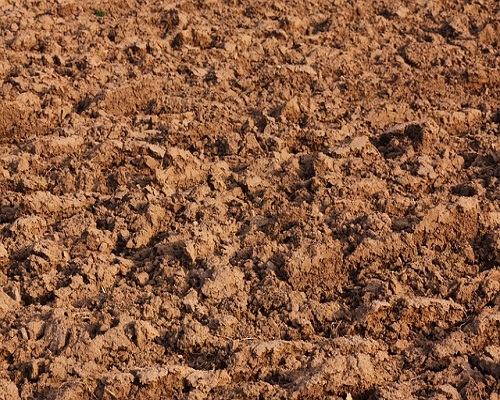}

\end{leftbar}

\subsection{Measures}\label{appx:measures}
This section presents the specific items used to measure each construct. For items whose content varies by the target emotion of a session, we use anger as the illustrative example, with corresponding adjustments made for fear sessions. We also use the group AI condition as the example. When the content differs for the single-AI condition, it is updated accordingly.

\subsubsection{Perceived Efficacy}

Looking back at the interactions with the AI system about the anger evoking issue, to what extent do you agree with the following statements?  \textit{(1=not at all, 7=very much)} \par

\begin{itemize}
    \item The interactions made me feel better.
    \item The interactions enabled me to take a different perspective on the issue that I discussed.
    \item The interactions ended up reducing my negative feelings.
\end{itemize}

\subsubsection{Connectedness}

Please click the button on the left to review your interactions with the AI agents about the anger-evoking issue. The review should be brief and help you recall your experience with each of the agents.
    
After this review, rate how connected you feel to each of the agents.\textit{(1=not at all, 7=very much)}
\begin{itemize}
    \item Agent 1
    \item Agent 2
\end{itemize}

\subsubsection{Trustworthiness}
To what extent do you agree with the following statements? \textit{(1=not at all, 7=very much)}
\begin{itemize}
    \item I think the agents in the AI system are trustworthy.
    \item I think the agents in the AI system are honest.
    \item I think the agents in the AI system are dependable.
\end{itemize}

\subsubsection{Expectation Disconfirmation}
Below are statements dealing with your perception of the AI system you’ve just interacted with regarding the anger evoking issue. Please indicate to what extent you agree with each statement. \textit{(1=not at all, 7=very much)}

\begin{itemize}
    \item The level of support provided by the AI system is what I would expect most AI systems to provide.
    \item The level of support provided by the AI system was exactly what I expected.
    \item Overall, most of my expectations regarding the level of support provided by the AI system were confirmed.
\end{itemize}

\section{Additional Results for Studies}

\setcounter{table}{0}
\renewcommand{\thetable}{B\arabic{table}}
\setcounter{figure}{0}
\renewcommand{\thefigure}{B\arabic{figure}}

\subsection{Additional Results for Study 1}
\begin{table}[!htbp]
    \renewcommand{\arraystretch}{0.5}
    \centering
    \caption{Manipulation Check of Emotion Condition in Study 1}
    \label{table:roomEmoManip}
    \begin{tabularx}{\textwidth}{
        >{\raggedright\arraybackslash}X
        >{\raggedright\arraybackslash}X
        >{\centering\arraybackslash}X
        >{\centering\arraybackslash}X
        >{\centering\arraybackslash}X
        >{\centering\arraybackslash}X
        >{\centering\arraybackslash}X
        >{\centering\arraybackslash}X
    }
    \midrule
    \multicolumn{2}{c}{Emotion} & \multicolumn{2}{c}{Mean} &
    \multirow{2}{*}{$t$} & \multirow{2}{*}{DoF} & \multirow{2}{*}{$p$} \\
    \cmidrule(lr){1-2} \cmidrule(lr){3-4}
    Target & Other & Target & Other &  &  &  \\ \midrule
        anger & disgust & 73.48  & 49.00  & 8.72  & 130  & \textless0.001 \\ 
        anger & fear & 73.48  & 18.88  & 17.45  & 130  & \textless0.001 \\ 
        anger & happiness & 73.48  & 8.53  & 26.95  & 130  & \textless0.001 \\ 
        anger & sadness & 73.48  & 48.61  & 7.66  & 130  & \textless0.001 \\ 
        anger & surprise & 73.48  & 30.30  & 14.16  & 130  & \textless0.001 \\ 
        fear & anger & 71.82  & 30.30  & 13.16  & 130  & \textless0.001 \\ 
        fear & disgust & 71.82  & 28.05  & 13.90  & 130  & \textless0.001 \\ 
        fear & happiness & 71.82  & 9.93  & 24.72  & 130  & \textless0.001 \\ 
        fear & sadness & 71.82  & 57.02  & 5.66  & 130  & \textless0.001 \\ 
        fear & surprise & 71.82  & 20.98  & 16.18  & 130  & \textless0.001 \\ \midrule

    \end{tabularx}
\end{table}

\begin{table}[!htbp]
    \renewcommand{\arraystretch}{0.5}
    \centering
    \caption{Balance Check of Study 1}
    \label{roomBalance}
    \begin{tabularx}{\textwidth}{
        >{\centering\arraybackslash}X
        >{\centering\arraybackslash}X
        >{\centering\arraybackslash}X
        >{\centering\arraybackslash}X
    }
    \midrule
        Variable & $\chi^2$ & $p$ & DoF \\ \midrule
        gender & 1.05  & 0.59  & 2 \\ 
        race & 5.40  & 0.49  & 6 \\
        education & 1.27  & 0.53  & 2 \\
        income & 9.90  & 0.27  & 8 \\ 
        Age & 0.25  & 0.88  & 2 \\ \midrule
    \end{tabularx}
\end{table}

\begin{figure}[!htbp]
\centering
\includegraphics[width=0.65\textwidth]{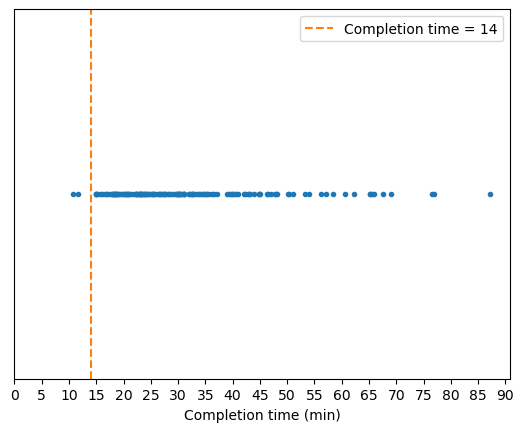}
\caption{Distribution of Completion Time}
\label{fig:completionDist}
\end{figure}

\clearpage
\subsection{Additional Results for Study 2}

\begin{table}[!htbp]
    \renewcommand{\arraystretch}{0.5}
    \centering
    \caption{Means and Standard Deviations of Pre-interaction Intensity of Emotions in Study 2}
    \label{table:strucEmoMean}
    \begin{tabularx}{\textwidth}{>{\arraybackslash}X>{\centering\arraybackslash}X>{\centering\arraybackslash}X}
    \midrule
        ~ & \multicolumn{2}{c}{Target Emotion of the Session} \\ \cmidrule(lr){2-3}
        ~ & Anger Session & Fear Session \\ 
        ~ & M (SD) & M (SD) \\ \midrule
        anger & 78.19 (20.31) & 41.48 (32.4) \\ 
        disgust & 57.02 (33.44) & 38.2 (34.19) \\ 
        fear & 23.79 (28.84) & 74.52 (23.7) \\ 
        happiness & 8.43 (11.6) & 10.37 (15.01) \\ 
        sadness & 61.5 (31.81) & 62.71 (30.0) \\ 
        surprise & 29.67 (33.64) & 21.78 (27.21) \\ \midrule
    \end{tabularx}
    \begin{minipage}{\textwidth}
        {\singlespacing \vspace{-\baselineskip}
        \footnotesize \textit{Notes:} M = mean, SD = standard deviation.}
    \end{minipage}
\end{table} 

\begin{table}[!htbp]
    \renewcommand{\arraystretch}{0.5}
    \centering
    \caption{Manipulation Check of Emotion Condition in Study 2}
    \label{table:strucEmoManip}
    \begin{tabularx}{\textwidth}{
        >{\raggedright\arraybackslash}X
        >{\raggedright\arraybackslash}X
        >{\centering\arraybackslash}X
        >{\centering\arraybackslash}X
        >{\centering\arraybackslash}X
        >{\centering\arraybackslash}X
        >{\centering\arraybackslash}X
        >{\centering\arraybackslash}X
    }
    \midrule
    \multicolumn{2}{c}{Emotion} & \multicolumn{2}{c}{Mean} &
    \multirow{2}{*}{$t$} & \multirow{2}{*}{DoF} & \multirow{2}{*}{$p$} \\
    \cmidrule(lr){1-2} \cmidrule(lr){3-4}
    Target & Other & Target & Other &  &  &  \\
    \midrule
    anger & disgust & 78.19  & 57.02  & 7.63  & 128  & \textless 0.001 \\ 
    anger & fear & 78.19  & 23.79  & 18.27  & 128  & \textless0.001 \\ 
    anger & happiness & 78.19  & 8.43  & 29.41  & 128  & \textless0.001 \\ 
    anger & sadness & 78.19  & 61.50  & 5.18  & 128  & \textless0.001 \\ 
    anger & surprise & 78.19  & 29.67  & 15.23  & 128  & \textless0.001 \\ 
    fear & anger & 74.52  & 41.48  & 10.33  & 126  & \textless0.001 \\ 
    fear & disgust & 74.52  & 38.21  & 10.96  & 126  & \textless0.001 \\ 
    fear & happiness & 74.52  & 10.37  & 25.46  & 126  & \textless0.001 \\ 
    fear & sadness & 74.52  & 62.71  & 4.60  & 126  & \textless0.001 \\ 
    fear & surprise & 74.52  & 21.78  & 18.47  & 126  & \textless0.001 \\ \midrule
    \end{tabularx}
\end{table}

\begin{table}[!htbp]
    \renewcommand{\arraystretch}{0.5}
    \centering
    \caption{Balance Check of Study 2}
    \label{strucBalance}
    \begin{tabularx}{\textwidth}{
        >{\centering\arraybackslash}X
        >{\centering\arraybackslash}X
        >{\centering\arraybackslash}X
        >{\centering\arraybackslash}X
    }
    \midrule
        Variable & $\chi^2$ & p-value & DoF \\ \midrule
        gender & 4.76  & 0.31  & 4 \\ 
        race & 12.37  & 0.42  & 12 \\ 
        education & 4.51  & 0.34  & 4 \\ 
        income & 21.67  & 0.15  & 16 \\ 
        Age & 4.43  & 0.35  & 4 \\ \midrule
    \end{tabularx}
\end{table}

\clearpage
\subsection{Additional Results for Study 3}
\begin{table}[!htbp]
    \renewcommand{\arraystretch}{0.5}
    \centering
    \caption{Means and Standard Deviations of Pre-interaction Intensity of Emotions in Study 3}
    \label{table:funcEmoMean}
    \begin{tabularx}{\textwidth}{>{\arraybackslash}X>{\centering\arraybackslash}X>{\centering\arraybackslash}X}
    \midrule
        ~ & \multicolumn{2}{c}{Target Emotion of the Session} \\ \cmidrule(lr){2-3}
        ~ & Anger Session & Fear Session \\ 
        ~ & M (SD) & M (SD) \\ \midrule
        anger & 72.28 (24.18) & 33.37 (31.19) \\ 
        disgust & 45.99 (33.41) & 24.99 (30.00) \\ 
        fear & 21.40 (28.73) & 67.85 (24.97) \\ 
        happiness & 8.04 (12.03) & 10.77 (14.78) \\ 
        sadness & 54.06 (33.76) & 53.25 (32.03) \\ 
        surprise & 25.25 (30.39) & 15.99 (23.23) \\ \midrule
    \end{tabularx}
    \begin{minipage}{\textwidth}
        {\singlespacing \vspace{-\baselineskip}
        \footnotesize \textit{Notes:} M = mean, SD = standard deviation. }
    \end{minipage}
\end{table}

\begin{table}[!htbp]
    \renewcommand{\arraystretch}{0.5}
    \centering
    \caption{Manipulation Check of Emotion Condition in Study 3}
    \label{table:funcEmoManip}
    \begin{tabularx}{\textwidth}{
        >{\raggedright\arraybackslash}X
        >{\raggedright\arraybackslash}X
        >{\raggedright\arraybackslash}X
        >{\raggedright\arraybackslash}X
        >{\raggedright\arraybackslash}X
        >{\raggedright\arraybackslash}X
        >{\raggedright\arraybackslash}X
        >{\raggedright\arraybackslash}X
    }
    \midrule
    \multicolumn{2}{c}{Emotion} & \multicolumn{2}{c}{Mean} &
    \multirow{2}{*}{$t$} & \multirow{2}{*}{DoF} & \multirow{2}{*}{$p$} \\
    \cmidrule(lr){1-2} \cmidrule(lr){3-4}
    Target & Other & Target & Other &  &  &  \\
    \midrule
    anger & disgust & 72.28  & 45.99  & 12.64  & 201.00  & \textless0.001 \\ 
        anger & fear & 72.28  & 21.40  & 19.35  & 201.00  & \textless0.001 \\ 
        anger & happiness & 72.28  & 8.05  & 32.66  & 201.00  & \textless0.001 \\
        anger & sadness & 72.28  & 54.06  & 6.79  & 201.00  & \textless0.001 \\ 
        anger & surprise & 72.28  & 25.25  & 18.77  & 201.00  & \textless0.001 \\
        fear & anger & 67.85  & 33.37  & 14.35  & 182.00  & \textless0.001 \\ 
        fear & disgust & 67.85  & 24.99  & 17.74  & 182.00  & \textless0.001 \\ 
        fear & happiness & 67.85  & 10.77  & 26.84  & 182.00  & \textless0.001 \\ 
        fear & sadness & 67.85  & 53.25  & 7.89  & 182.00  & \textless0.001 \\
        fear & surprise & 67.85  & 16.00  & 23.25  & 182.00  & \textless0.001 \\ \midrule 
    \end{tabularx}
\end{table}

\begin{table}[!htbp]
    \renewcommand{\arraystretch}{0.5}
    \newcommand{\tightcell}[1]{\begin{tabular}[c]{@{}c@{}}\linespread{1}\selectfont #1\end{tabular}}
    \centering
    \caption{Manipulation Check of Perceived Support Type of Agents}
    \label{table:funcSupportCheck}
    \centering
    \begin{tabularx}{\textwidth}{
        >{\raggedright\arraybackslash}X
        >{\raggedright\arraybackslash}X
        >{\raggedright\arraybackslash}X
        >{\centering\arraybackslash}X
        >{\centering\arraybackslash}X
        >{\centering\arraybackslash}X
        >{\centering\arraybackslash}X
        >{\centering\arraybackslash}X
    }
    \midrule
        Group & Agent & \tightcell{Expected \\Type} & \tightcell{Mean\\(Emotional)} & \tightcell{Mean\\(Informational)} & $t$ & DoF & $p$ \\ \midrule
        Base & agent 1 & $\backslash$ & 5.64  & 5.54  & 0.73  & 73 & 0.467  \\ 
        Base & agent 2 & $\backslash$ & 5.82  & 5.69  & 1.11  & 73 & 0.272  \\ \midrule
        Emo\_Info& agent 1 & Emotional & 6.08  & 4.70  & 6.08  & 82 & \textless0.001  \\ 
        Emo\_Info& agent 2 & Informational & 4.45  & 5.96  & -7.56  & 82 & \textless0.001  \\ \midrule
        Info\_Emo & agent 1 & Informational & 4.64  & 5.81  & -6.76  & 82 & \textless0.001  \\ 
        Info\_Emo & agent 2 & Emotional & 6.05  & 4.39  & 8.99  & 82 & \textless0.001  \\ \midrule
        2Emo & agent 1 & Emotional & 5.40  & 4.53  & 4.19  & 67 & \textless0.001  \\ 
        2Emo & agent 2 & Emotional & 5.43  & 4.72  & 3.47  & 67 & 0.001  \\ \midrule
        2Info & agent 1 & Informational & 4.47  & 5.10  & -3.53  & 76 & 0.001  \\ 
        2Info & agent 2 & Informational & 4.38  & 5.06  & -4.11  & 76 & \textless0.001  \\ \midrule
    \end{tabularx}
\end{table}

\begin{table}[!htbp]
    \renewcommand{\arraystretch}{0.5}
    \centering
    \caption{Balance Check of Study 3}
    \label{table:funcBalance}
    \begin{tabularx}{\textwidth}{
        >{\centering\arraybackslash}X
        >{\centering\arraybackslash}X
        >{\centering\arraybackslash}X
        >{\centering\arraybackslash}X
    }
    \midrule
        Variable & $\chi^2$ & p-value & DoF \\ \midrule
        gender & 1.37  & 0.85  & 4 \\
        race & 32.41  & 0.26  & 28 \\
        education & 2.79  & 0.95  & 8 \\
        income & 32.33  & 0.45  & 32 \\
        Age & 10.01  & 0.26  & 8 \\ \midrule
    \end{tabularx}
\end{table}

\clearpage
\subsection{Additional Results for the Moderated Mediation Analysis}
\begin{table}[!htbp]
    \renewcommand{\arraystretch}{0.5}
    \centering
    \caption{Conditional Indirect Effect of Group AI Support on Perceived Efficacy by Income Group}
    \label{table:modCondEff}
    \begin{tabularx}{\textwidth}{
        >{\raggedright\arraybackslash}X
        >{\centering\arraybackslash}X
        >{\centering\arraybackslash}X
        >{\centering\arraybackslash}X
        >{\centering\arraybackslash}X
    }
    \midrule
        Income Level & $\beta$ & SE & 90\% CI & Zero Included? \\ \midrule
        Low Income & 0.94 & 0.27 & [0.50, 1.37] & No \\ 
        Not Low Income & 0.27 & 0.16 & [0.01, 0.53] & No \\ \midrule
    \end{tabularx}
    \begin{minipage}{\textwidth}
        {\singlespacing \vspace{-\baselineskip}
        \footnotesize \textit{Notes:} SE = standard error, CI = confidence interval. For the "Zero Included?" column, “No” indicates that zero is not contained within the reported confidence interval, which suggests evidence of a nonzero indirect effect. \par}
    \end{minipage}
\end{table}

\clearpage
\section{Pilot Study: An Online Forum–Style Interface} \label{appx:forumStyle}

\setcounter{table}{0}
\renewcommand{\thetable}{C\arabic{table}}
\setcounter{figure}{0}
\renewcommand{\thefigure}{C\arabic{figure}}

\begin{figure}[!htbp]
    \centering
    \subfloat[\small Group AI (Forum Interface)]{\label{fig:forumPageGroup} \includegraphics[width=.8\linewidth]{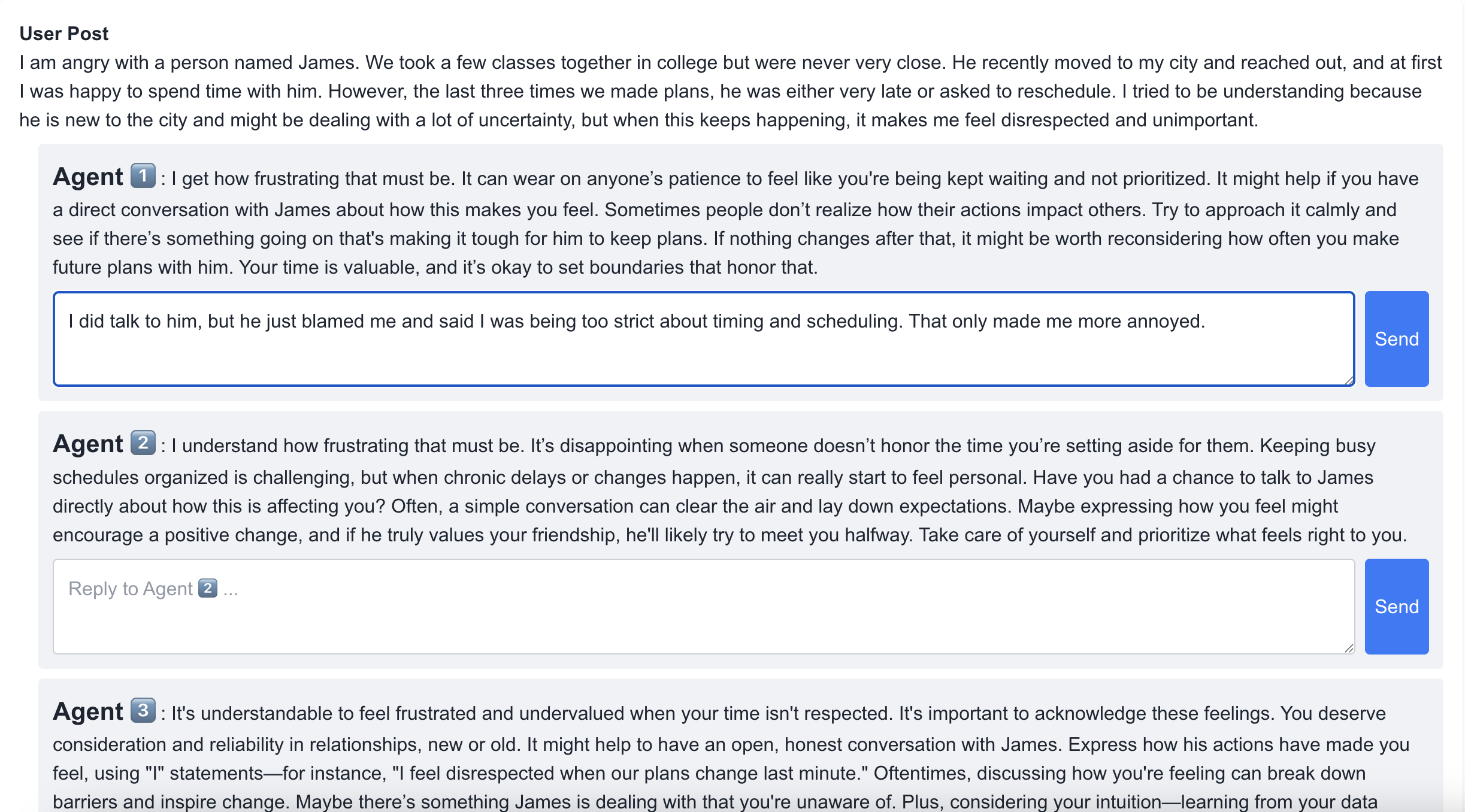}}\par
    \subfloat[\small Single AI (Forum Interface)]{\label{fig:forumPagesingle} \includegraphics[width=.8\linewidth]{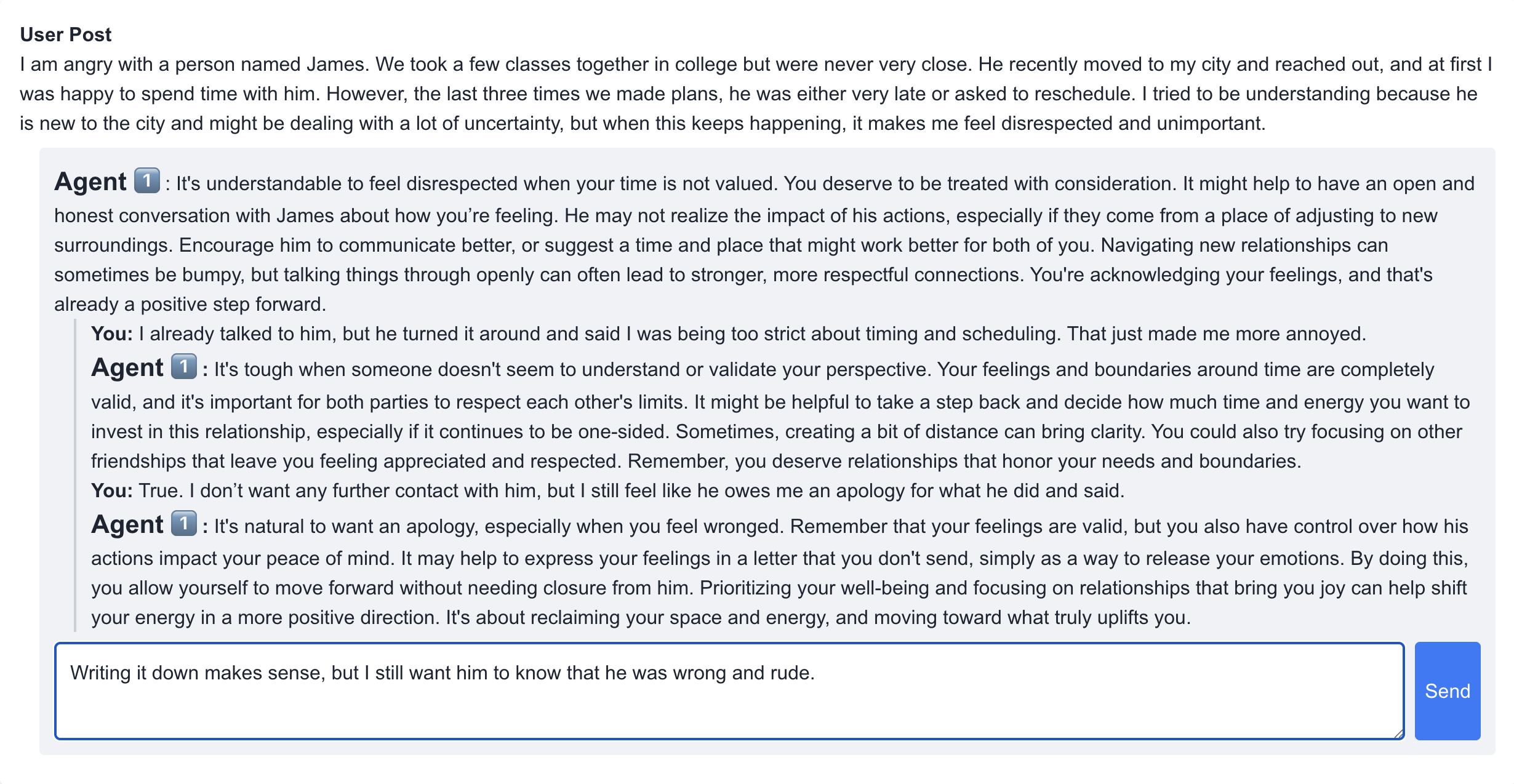}}\\ 
    \caption{Forum-Style Interface} \label{fig:forumPage}
\end{figure}

Before conducting the formal studies, we ran a pilot study using an online forum–style interface in which participants interacted with multiple AI agents (see Figure \ref{fig:forumPage}). We recruited 80 participants (34 male, 45 female) from Prolific. Apart from the interface design, participants followed the same experimental procedure, and all constructs were measured using the same questionnaire described in Section \ref {sec:study1} of the manuscript.

In the group AI condition of the forum-style interface, participants interacted with four AI agents, all powered by GPT-4o. To reduce unintended repetition across agents that could interfere with the results, we first used a single API call to generate four distinct personas. Each persona specified major demographic information, such as age, gender and a brief occupation description. These personas were then assigned to the four agents, and separate API calls were used to generate each agent’s responses and sustain the interaction with participants. Personas were not used in the single AI condition, as there was no risk of overlap across agents. Across both conditions, all AI agents were prompted with the same overarching goal of alleviating users’ negative emotions.

We analyzed data from 74 participants who satisfied all data screening criteria specified in Section \ref{sec:study1} of the manuscript, among whom 58 met these criteria in both the anger and fear sessions. Prior to interacting with the AI system, participants reported higher intensity for the target emotion (anger or fear) than for non-target emotions in the corresponding session, indicating successful emotion induction. Relevant results are reported in Tables \ref{table:forumEmoMean} and \ref{table:forumEmoManip}. To examine the effect of group AI versus single AI support in the forum-style interface on perceived support efficacy, we conducted an ANOVA with support condition as a between-subjects factor. The results indicate that, under the forum-style interface, the single AI condition led to higher perceived support efficacy than the group AI condition ($M_{group}$ = 4.986 versus $M_{single}$ = 5.439, $SD$s= 1.407 and 1.349, $F$(1,130) = 3.561, $p$ = 0.061).

\begin{table}[!htbp]
    \renewcommand{\arraystretch}{0.5}
    \centering
    \caption{Means and Standard Deviations of Pre-interaction Intensity of Emotions in \\the Forum-Style Pilot Study}
    \label{table:forumEmoMean}
    \begin{tabularx}{\textwidth}{>{\arraybackslash}X>{\centering\arraybackslash}X>{\centering\arraybackslash}X}
    \midrule
        ~ & \multicolumn{2}{c}{Target Emotion of the Session} \\ \cmidrule(lr){2-3}
        ~ & Anger Session & Fear Session \\ 
        ~ & M (SD) & M (SD) \\ \midrule
        anger & 75.56 (20.71) & 26.10 (25.55) \\ 
        disgust & 50.40 (34.31) & 25.86 (29.33) \\
        fear & 17.40 (23.88) & 74.37 (21.36) \\
        happiness & 8.98 (15.82) & 10.03 (14.79) \\
        sadness & 52.92 (32.89) & 56.86 (28.97) \\
        surprise & 27.05 (30.2) & 20.50 (25.9) \\ \midrule
    \end{tabularx}
    \begin{minipage}{\textwidth}
        {\singlespacing \vspace{-\baselineskip}
        \footnotesize \textit{Notes:} M = mean, SD = standard deviation. }
    \end{minipage}
\end{table}

\begin{table}[!htbp]
    \renewcommand{\arraystretch}{0.5}
    \centering
    \caption{Manipulation Check of Emotion Condition in the Forum-Style Pilot Study}
    \label{table:forumEmoManip}
    \begin{tabularx}{\textwidth}{
        >{\raggedright\arraybackslash}X
        >{\raggedright\arraybackslash}X
        >{\raggedright\arraybackslash}X
        >{\raggedright\arraybackslash}X
        >{\raggedright\arraybackslash}X
        >{\raggedright\arraybackslash}X
        >{\raggedright\arraybackslash}X
        >{\raggedright\arraybackslash}X
    }
    \midrule
    \multicolumn{2}{c}{Emotion} & \multicolumn{2}{c}{Mean} &
    \multirow{2}{*}{$t$} & \multirow{2}{*}{DoF} & \multirow{2}{*}{$p$} \\
    \cmidrule(lr){1-2} \cmidrule(lr){3-4}
    Target & Other & Target & Other &  &  &  \\
    \midrule
        anger & disgust & 75.56  & 50.40  & 5.65  & 61  & \textless0.001 \\ 
        anger & fear & 75.56  & 17.40  & 16.01  & 61  & \textless0.001 \\ 
        anger & happiness & 75.56  & 8.98  & 20.53  & 61  & \textless0.001 \\ 
        anger & sadness & 75.56  & 52.92  & 4.49  & 61  & \textless0.001 \\ 
        anger & surprise & 75.56  & 27.05  & 10.70  & 61  & \textless0.001 \\ 
        fear & anger & 74.37  & 26.10  & 12.92  & 69  & \textless0.001 \\ 
        fear & disgust & 74.37  & 25.86  & 11.85  & 69  & \textless0.001 \\ 
        fear & happiness & 74.37  & 10.03  & 19.95  & 69  & \textless0.001 \\ 
        fear & sadness & 74.37  & 56.86  & 5.87  & 69  & \textless0.001 \\ 
        fear & surprise & 74.37  & 20.50  & 14.84  & 69  & \textless0.001 \\ \midrule
    \end{tabularx}
\end{table}

These findings contrast with the results reported in Section \ref{sec:study1} of the manuscript. When using a forum-style interface, group AI support did not provide a more effective support form and instead appeared to be less effective than single AI support. One possible explanation is that the forum-style interface organizes each agent’s response into a separate thread, which may lead users to experience the interaction as maintaining multiple parallel conversations rather than receiving support from a coherent group. As a result, relative to the mechanisms discussed in Section \ref{sec:hypothesis} of the manuscript, ties among agents are not salient, and the coexistence of direct and indirect ties that could strengthen relational perceptions may fail to emerge. The psychological strain is also unlikely to arise. Moreover, switching across multiple threads may hinder sustained engagement within any single thread, further impeding the development of connection between the user and individual agents. These results highlight interface design as an important consideration for group AI support applications and suggest that an online forum–style interface may represent a suboptimal design choice in practice.
\end{appendices}

\clearpage

\end{document}